\documentclass[]{siamart171218}

\usepackage[utf8x]{inputenc}
\usepackage[english]{babel}

\usepackage{amsmath}
\usepackage{amsfonts}
\usepackage{amssymb}
\usepackage{bm}

\usepackage{graphicx}
\usepackage{cite}
\usepackage{color}
\usepackage{standalone}
\usepackage{tabularx,booktabs}
\usepackage[margin=0pt,font+={small,it},labelfont+={sc},labelsep=space,list=false,hypcap=false,skip=3pt,belowskip=-12pt]{caption}
\usepackage[skip=0pt]{subcaption}
\usepackage{microtype}

\usepackage{tikz}
\usetikzlibrary{patterns}
\usetikzlibrary{arrows}
\usetikzlibrary{decorations.markings}
\usetikzlibrary{cd}
\usetikzlibrary{positioning,backgrounds,calc}

\usepackage[normalem]{ulem}

\newcommand{\SC}{\ensuremath{\mathcal{X}}}
\newcommand{\simplex}[1]{\ensuremath{\mathcal{S}^{#1}}}
\newcommand{\1}[0]{\ensuremath{\bm{1}}}
\newcommand{\nLap}[0]{\ensuremath{\bm{\mathcal{L}}}}
\newcommand{\liftB}[0]{\ensuremath{\widehat{\bm{B}}}}

\newtheorem{example}{Example}

\newcommand{\phantomsubfigure}[1]{\begin{subfigure}[b]{0.1\textwidth}\phantomcaption\label{#1}\end{subfigure}}
\definecolor{mylinkcolor}{RGB}{0,0,140}
\hypersetup{colorlinks,allcolors=mylinkcolor,citecolor=mylinkcolor}

\begin{document}
\title{Random walks on simplicial complexes and the normalized Hodge 1-Laplacian\thanks{
        Submitted to the editors DATE.
\funding{
MTS acknowledges funding from the European Union's Horizon 2020 research and innovation programme under the Marie Sklodowska-Curie grant agreement No 702410. 
ARB was supported in part by NSF Award DMS-1830274.
GL, PH, and AJ acknowledge funding from AFOSR Grant FA9550-13-1-0097.
AJ's research was supported in parts by a Vannevar Bush Faculty Fellowship from the Office of Secretary of Defense.
ARB and AJ were supported in part by ARO Award W911NF-19-1-0057.
The funders had no role in the design of this study; the results presented here reflect solely the authors' views.}}}
\author{Michael T. Schaub\thanks{Institute for Data, Systems and Society, Massachusetts Institute of Technology\& Department of Engineering Science, University of Oxford. \email{mschaub@mit.edu}}
    \and Austin R. Benson\thanks{Department of Computer Science, Cornell University \email{arb@cs.cornell.edu}}
    \and Paul Horn\thanks{Department of Mathematics, University of Denver \email{paul.horn@du.edu}}
    \and Gabor Lippner\thanks{Department of Mathematics, Northeastern University \email{g.lippner@neu.edu}}
    \and Ali Jadbabaie\thanks{Institute for Data, Systems and Society, MIT \email{jadbabai@mit.edu}}
}
\maketitle

\begin{abstract}
Using graphs to model pairwise relationships between entities is a ubiquitous framework for studying complex systems and data.
Simplicial complexes extend this dyadic model of graphs to polyadic relationships and have emerged as a model for multi-node relationships occurring in many complex systems.
For instance, biological interactions occur between sets of molecules and communication systems include group messages that are not pairwise interactions.
While Laplacian dynamics have been intensely studied for graphs, corresponding notions of Laplacian dynamics beyond the node-space have so far remained largely unexplored for simplicial complexes.
In particular, diffusion processes such as random walks and their relationship to the graph Laplacian---which underpin many methods of network analysis, including centrality measures, community detection, and contagion models---lack a proper correspondence for general simplicial complexes.

Focusing on coupling between edges, we generalize the relationship between the normalized graph Laplacian and random walks on graphs by devising an appropriate normalization for the Hodge Laplacian---the generalization of the graph Laplacian for simplicial complexes---and relate this to a random walk on edges.
Importantly, these random walks are intimately connected to the topology of the simplicial complex, just as random walks on graphs are related to the topology of the graph.
This serves as a foundational step towards incorporating Laplacian-based analytics for higher-order interactions.
We demonstrate how to use these dynamics for data analytics that extract information about the edge-space of a simplicial complex that complements and extends graph-based analysis.
Specifically, we use our normalized Hodge Laplacian to derive spectral embeddings for examining trajectory data of ocean drifters near Madagascar and also develop a generalization of personalized PageRank for the edge-space of simplicial complexes to analyze a book co-purchasing dataset.
\end{abstract}
 

\section{Introduction}
Markov chains and diffusion are staples of applied mathematics~\cite{Billingsley-1961-statistical,Ikeda-2014-stochastic,Masuda2017,Boyd-2004-fastest}.
Any time-homogeneous finite state Markov chain can be interpreted as a random walk on a graph: the states of the Markov chain are the nodes of the graph, and transitions occur between connected nodes via an appropriately weighted edge.
This close relationship between Markov chains and graphs has led to a broad adoption of diffusion-based algorithms in network science~\cite{Masuda2017}, with applications to ranking connected objects~\cite{Gleich2015}, analyzing disease spreading~\cite{Salath2010}, and respondent-driven sampling~\cite{Salganik2004}.
Key to the success of many such algorithms is the link between random walks and (the spectral theory of) the graph Laplacian~\cite{Masuda2017,Chung1997,Chung2006}, a matrix that encodes the structure of a graph and has intimate connections to discrete potential theory and harmonic analysis~\cite{biggs1997algebraic}.
Indeed, there is well-developed theory relating topological properties of graphs to features of the graph Laplacian and thus to random walks and diffusion processes~\cite{Chung1997,Chung2006,biggs1997algebraic}.
For instance, spectral properties of the graph Laplacian are related to expansion, diameter, distance between subsets, and the mixing time of random walks, amongst others~\cite{Chung1997,Chung2006}.
Thus, analyzing the properties of a random walk on a network (i.e., graph), or alternatively the graph Laplacian, can reveal fundamental properties about the system under investigation.

As network-based system models have become almost ubiquitous across scientific domains~\cite{newman2003structure,boccaletti2006complex}, 
graphs and their Laplacians feature prominently in many analysis tasks~\cite{Coifman2006,belkin2002laplacian,Masuda2017,Guattery1998,mohar1991laplacian,Benson2016}.
However, graphs are in fact special cases of more general mathematical objects called simplicial complexes (SCs),\footnote{Formally, we use \emph{abstract simplicial complexes}, but we drop ``abstract''
  for easier reading.} 
and the graph Laplacian is a special case of the Hodge Laplacians that appears in algebraic topology~\cite{Lim-2015-hodge}.
We give a formal definition of SCs later, but an intuitive description of an SC is a generalization of a graph whose edge-set can contain subsets of any number of nodes (more specifically, a hypergraph with certain properties).
Since SCs can describe richer sets of relationships than a graph, SCs are increasingly used to analyze systems and data (see \Cref{sec:motivation_background}).

In this paper, we introduce a certain normalized Hodge Laplacian matrix and show how it relates to random walks on SCs, 
with an overarching goal to develop data analyses that respect additional aspects of the topology of the data.
A primary motivation for this study of diffusion on SCs is to facilitate the translation of the large toolbox of network science for graphs to simplicial complexes.
To make our results concrete, we present our work in the language of linear algebra and focus on devising random walks on ``1-simplices,'' which may be thought of as edges.
Stated differently, our results provide tools for the analysis of signals defined in the \emph{edge-space} of a SC (or graph) that complement the typical node-based analysis conducted for graphs.
In particular, we present tools that enable us to extract the relative importance of edges and edge-signals with respect to the higher-order topological properties of the SC.
We contrast our methodology with other notions of edge-based random walks such as those based on line graphs~\cite{Evans2009,Ahn2010} and consider how higher-order interactions leads to certain difficulties in formulating a diffusion model absent in the theory of random walk on graphs~\cite{Mukherjee2016,Rosenthal2014}.

We show two applications to illustrate our ideas, which show how our methodology incorporates higher-order topology of the data into the analysis.

In our first application, we develop embeddings of edge-flows and trajectory data as a higher-order generalization of diffusion maps~\cite{Coifman2006} and Laplacian eigenmaps~\cite{belkin2002laplacian}.
Similar to the embedding of the nodes of a graph into a Euclidean space, this embedding provides us with an effective low-dimensional representation of edges and trajectories in a vector space.
This vector space representation can then be used to gain further insights about the observed flows.
Here we illustrate how the embedding can be used to differentiate types of trajectories, but other data analysis and machine learning tasks may be addressed with these techniques as well.

Our second application is a variant of (personalized) PageRank~\cite{Gleich2015} for 1-simplices (edges in a graph).
Here we demonstrate how our tools enable us to analyze the ``role'' certain edges play with respect to the global topology of the SC.
We point out how these tools may be seen as extensions of ideas from graph signal processing, typically concerned with signals on nodes, to the space of signals defined on edges.
Indeed, we demonstrate how our analysis is complementary to node-based analysis and how our tools can highlight, e.g., how much an edge is part of the cycle-space of the SC.

\subsection{Additional Background and Motivation}\label{sec:motivation_background}
\paragraph{Networks as models for complex systems}
Many complex systems are naturally modeled by
graphs~\cite{newman2003structure}. Due to the broad scope of this modeling
paradigm, the analysis of systems as networks by means of graph-theoretic tools
has been tremendously successful, with applications in biological,
technological, and social
systems~\cite{boccaletti2006complex,Newman2010,newman2003structure,Strogatz2001,albert2002statistical}.
However, graph-based representations of complex systems have some limitations.
Specifically, graphs encode pairwise relationships between entities via edges,
but do not account for (simultaneous) interactions between more than two nodes,
which can be crucial for the ensuing dynamics of a
system~\cite{Benson2016,grilli2017higher}.
For instance, biochemical reactions often involve more than two species in a
reaction~\cite{Klamt2009}; sensors record collections of interactions at a given
time~\cite{Ghrist2005,DeSilva2007,Tahbaz-Salehi2010}; people communicate in
small groups with group chat messaging; and individuals form teams or exert peer
pressure~\cite{Bonacich2004,Kee2013}.  Non-dyadic interactions have in fact long
been an object of study in the social sciences.  For instance, structural
balance theory implies that 3-way relationships in social networks will evolve
according to colloquial rules such as ``the friend of a friend is my friend''
and ``the enemy of my enemy is my friend''~\cite{Cartwright1956,Wasserman1994,Marvel2011}.

\paragraph{Modeling higher-order interactions}
There are several modeling frameworks for non-dyadic, higher-order interactions,
such as simplicial complexes~\cite{Hatcher2002,Lim-2015-hodge}, set
systems~\cite{Berge1973}, hypergraphs~\cite{Bollobas1986}, and affiliation
graphs~\cite{Feld1981}.
Here we focus on simplicial complexes, which, in contrast to generic
hypergraphs, have special algebraic structure that makes them a core object of
study in applied (algebraic)
topology~\cite{carlsson2009topology,ghrist2008barcodes}.
Such algebraic structure is accompanied by analogs of the
graph Laplacian for simplicial complexes, namely the Hodge Laplacian, which will
be a principal object of our study.

\paragraph{Related work} 
Simplicial complex models have been successful in gaining new insights into
biological data~\cite{Nanda2014,Chan2013},
structural and functional connectivity in the brain~\cite{Petri2014,Giusti2015,Giusti-2016-neural},
coverage of sensor networks~\cite{Ghrist2005,DeSilva2007,Tahbaz-Salehi2010,Muhammad2006},
signal processing~\cite{barbarossa2016introduction,schaub2018flow},
mobility analysis~\cite{ghosh2018topological},
and robotics~\cite{Pokorny216}.
Simplicial models have further been studied from a geometric
perspective~\cite{bianconi2017emergent}, in terms of epidemic
spreading~\cite{iacopini2018simplicial}, or in the context of extensions of
random graph
models~\cite{kahle2014topology,courtney2016generalized,young2017construction}.
Nevertheless, in contrast to graph-based methods, the analysis of higher-order
interaction data using simplicial complexes is still nascent, even though the
formal use of tools from algebraic topology for the analysis of networks was
discussed already in the 1950s in the context of electrical network and circuit
theory~\cite{roth1955application,reza1958some,roth1959application}.
And Eckmann's seminal work introduced the ideas underpinning the Hodge Laplacian
already in 1944~\cite{Eckmann1944}.
That being said, little is known about the spectral properties of Hodge
Laplacians and how they relate to dynamics on the underlying simplicial
complexes.
Specifically, notions such as random walks and diffusion processes on simplicial
complexes have remained scarcely explored and mainly from the perspective of
pure mathematics~\cite{Parzanchevski2017,Rosenthal2014,Mukherjee2016,Horak2013}.
Preliminary research elucidating spectral connections include
spectral sparsification of simplicial complexes~\cite{Osting2017},
embeddedness of edges in the cycle-space of a graph~\cite{Schaub2014},
the analysis of flows on graphs and discretized domains~\cite{schaub2018flow,ghosh2018topological},
and the spectral theory of hypermatrices, tensors and hypergraphs~\cite{Benson2015,Benson2016,Gautier2018,Qi2017,Chan2018,Zhou2007}.

\subsection{Outline and Notation}
We first briefly review simplicial complexes and Hodge Laplacians in
\Cref{sec:background}. Then, in \Cref{sec:diffusion_on_SC}, we discuss our
normalized variant of the Hodge Laplacian, how it can be related to models for
diffusion processes on SCs in the edge-space, and analyze its spectral
properties.  \Cref{sec:computation_and_constructions} describes how SCs can be
constructed from data and discusses computational aspects of our formalism.
Finally, \Cref{sec:appplication_embedding,sec:applicationPR} outline trajectory
embeddings and simplicial PageRank as two applications of our random walk model on
SCs.

\paragraph*{Notation} 
Matrices and vectors are denoted by bold-faced fonts ($\bm{A}, \bm{x}$).
Their entries will be denoted by indexed letters (such as $A_{i,j}, x_i$), or for clarity as $(\bm A)_{i,j}$.
All vectors are assumed to be column vectors.
Scalar quantities are denoted by lowercase letters such as $a, b$.
We use $\1$ to denote the vector of all ones, and $\bm{I}$ to denote the identity matrix.
We use $|\bm{A}|$ to denote the matrix whose elements are given by the absolute values of the entries of $\bm A$.
We further denote the positive and negative parts of a real-valued matrix by $\bm{A}^+ := \max(\bm{A},\bm{0})$ and $\bm{A}^- := \max (-\bm{A},\bm{0})$, where the maximum is applied element-wise.
For a vector $\bm x$, the matrix $\text{diag}(\bm x)$ denotes the diagonal matrix whose (diagonal) entries are given by the components of the vector $\bm x$.
For square matrices $\bm A, \bm B$, we use $\text{diag}(\bm A,\bm B)$ to denote a block-diagonal matrix in which the matrices $\bm A, \bm B$ form the diagonal blocks (and the remaining entries are zero).
Sets are denoted by calligraphic letters such as $\mathcal S$, except for the real numbers, which we denote by $\mathbb{R}$.

\section{A short review of graphs, simplicial complexes, and Laplacians}\label{sec:background}
We briefly review some ideas from graph theory and algebraic topology.  Our
exposition is geared towards readers with an understanding of graphs and
matrices and is similar to the more detailed exposition by Lim~\cite{Lim-2015-hodge}.

\subsection{Graphs and the graph Laplacian}\label{sec:graphs_and_laplacians}
An undirected graph $\mathcal G$ consists of a set of vertices $\mathcal V$ with cardinality $\lvert \mathcal V \rvert =n_0$ and a set of edges $\mathcal E$, where each edge is an unordered pair of nodes.
For convenience, we identify the nodes with the integers $1, \ldots, n_0$.
The structure of a graph can be encoded in an adjacency matrix $\bm{A}$ with entries $A_{i,j} = 1$, if $i$ is connected to $j$ via an edge, and $A_{i,j} = 0$ otherwise.
As we consider undirected graphs here, $\bm{A} = \bm{A}^\top$.
A connected component is a maximal set of nodes $\mathcal{V}_c$ such that there exists a sequence of edges in the graph via which every node in $\mathcal{V}_c$ can be reached from every other node in $\mathcal{V}_c$.
The degree of a node $i$ is the number of edges containing $i$.
Accordingly, we can define the matrix of degrees as $\bm{D} := \text{diag}(\bm{A}\bm{1})$,
where $\text{diag}(\bm{x})$ is the diagonal matrix with the entries of $\bm{x}$ on the diagonal.
The graph Laplacian matrix $\bm{L}_0 = \bm{D} - \bm{A}$ is an algebraic
description of a graph, whose spectral properties reveal a number of important
topological properties of the
graph~\cite{mohar1991laplacian,spielman2007spectral,Bollobas2013,Chung1997,Chung2006}.

\subsection{Simplicial Complexes}
Let $\mathcal V$ be a finite set of vertices.
A \emph{k-simplex} $\simplex{k}$ is a subset of $\mathcal V$ of cardinality $k+1$ (we do not allow $\simplex{k}$ to be a multi-set, i.e., there are no repeated elements in $\simplex{k}$).
A simplicial complex (SC) $\SC$ is a set of simplices with the property that if $\simplex{} \in \SC$, then all subsets of $\simplex{}$ are also in $\SC$.

\begin{example}\label{Ex:1}
    Analogous to a graph, the vertices of $\mathcal X$ in \cref{fig:schematic_A} correspond to ``nodes'' $\{1,\ldots,7\}$ 
    and the {$1$-simplices} to ``edges.''
    The $2$-simplices $\{1,2,3\}$ and $\{2,3,4\}$ are depicted by filled triangles.
\end{example}

\begin{figure}[tb!]
\phantomsubfigure{fig:schematic_A}
\phantomsubfigure{fig:schematic_B}
    \centering
    \includegraphics{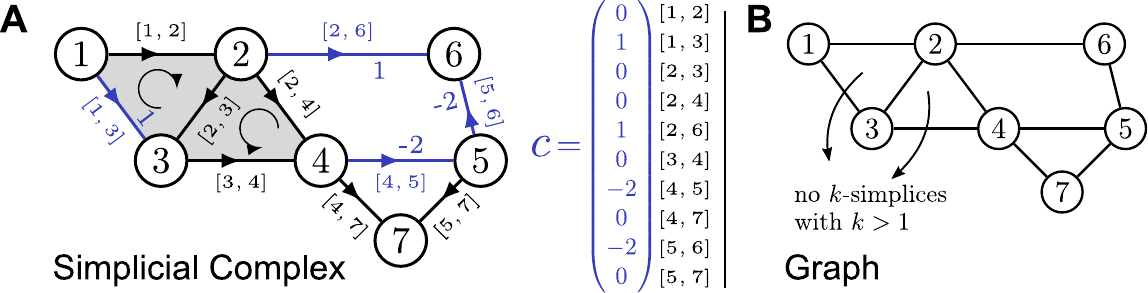}
    \caption{\textbf{Simplicial complexes and graphs}. 
    \textbf{A} Schematic of a simplicial complex with a prescribed orientation.
    This is the running example of a simplicial complex in the text. 
    Shaded areas correspond to the 2-simplices $\{1,2,3\}$ and $\{2,3,4\}$.
    An edge flow $c$ along the paths $2\rightarrow 6\rightarrow 5\rightarrow 4$ and $1 \rightarrow 3$ as well as
    its corresponding vector representation are depicted in blue. 
    \textbf{B} Schematic of a graph, corresponding to the $1$-skeleton of the simplicial complex in (A). 
    There are no $k$-simplices with $k>1$ in the graph.}    \label{fig:schematic}
\end{figure}

A graph, while typically defined via two sets (vertices and edges), 
may be interpreted as an SC in which all simplices have cardinality at most $2$ (\cref{fig:schematic_B}).
An SC can thus be understood as a generalization of a graph encoding higher-order relationships between vertices. 
To emphasize this connection, we will call the collection of $1$-simplices in an SC $\SC$ the edges of $\SC$.

A \emph{face} of a simplex $\simplex{k}$ is a subset of $\simplex{k}$ with cardinality $k$, i.e., with one element of $\simplex{k}$ omitted.
If $\simplex{k-1}$ is a face of simplex $\simplex{k}$, $\simplex{k}$ is called a \emph{co-face} of $\simplex{k-1}$.
\setcounter{example}{0}
\begin{example}[continued]
    In \cref{fig:schematic_A}, $\{1,2\}, \{2,3\}$ and $\{1,3\}$ are faces of $\{1,2,3\}$. Similarly $\{2,3,4\}$ is a co-face of $\{2,3\}$, $\{3,4\}$ and $\{2,4\}$.
\end{example}

Two $k$-simplices in an SC $\SC$ are \emph{upper adjacent}
if they are both faces of the same $(k+1)$-simplex and
are \emph{lower adjacent} if both have a common face.
For any $\simplex{}\subset \SC$, we define its degree, 
denoted by $\text{deg(\simplex{})}$, to be the number of co-faces of $\simplex{}$.
We use $\SC^{k}$ to denote the subset of $k$-simplices in $\SC$.
\setcounter{example}{0}
\begin{example}[continued]
    In \cref{fig:schematic_A}, the simplices $\{3,4\}$ and $\{2,4\}$ are upper adjacent, but $\{2,6\}$ and $\{5,6\}$ are not.
    The simplices $\{3,4\}$ and $\{2,4\}$ are lower adjacent, as are $\{2,6\}$ and $\{5,6\}$.
\end{example}

\subsection{Oriented sipmlicial complexes and function spaces on simplicial complexes}
While the definition of SCs is based on sets, in order to facilitate computations, 
we need to define an orientation for each simplex, which we do by fixing an ordering of its vertices.
The choice of orientation is a matter for book-keeping: just like we need to define a node-labeling to represent a graph with an adjacency matrix,
we need to define orientations to perform appropriate numerical computations for simplicial complexes.\footnote{For the expert, there is a small subtlety here.
  Strictly speaking, orientations are essential because we are concerned with signals defined on simplicial complexes represented by vectors with field coefficients in $\mathbb{R}$.
  If we were to use (binary) field coefficients in $\mathbb{Z}/2$, we would not have to define orientations. 
  However, there is no relevant Hodge theory associated to this case.
  For applications to data, real-valued coefficients are essential to represent signals on edges and will be our focus.}

Formally, an orientation of a $k$-simplex $\simplex{k}$ ($k>0$) is an equivalence class of orderings of its vertices, where two orderings are equivalent if they differ by an even permutation.
For simplicity, we choose the reference orientation of the simplices induced by the ordering of the vertex labels 
$\{[i_0,\ldots,i_k] : i_0 < \ldots < i_k\}$.

\setcounter{example}{0}
\begin{example}[continued]
In \cref{fig:schematic_A}, edges and triangles are oriented by arrows on the simplicies.
In this example, the ordered simplices $[2,3,4]$ and $[3,4,2]$ correspond to an equivalent orientation, whereas $[1,2]$ and $[2,1]$ do not.
\end{example}

A node (0-simplex) can have only one orientation.
Hence, issues of orientation do not commonly arise in graph-theoretic settings.
An exception are graph-flow problems, in which orientations are defined for edges as above to keep track of the flows appropriately: each flow has a magnitude and a sign to indicate if the direction of the flow is aligned or anti-aligned with the chosen reference orientation.

Given a reference orientation for each simplex, for each $k$, we can define the finite-dimensional vector space 
$\mathcal C_k$ with coefficients in $\mathbb R$, whose basis elements are the oriented simplices $s^k_i$.
An element $c_k \in \mathcal C_k$ is called a \emph{$k$-chain}, and may be thought of as a formal linear combination of these basis elements $c_k = \sum_i\gamma_i s^k_i$.
Thus, we can represent each element in $\mathcal C_k$ by a vector $\bm{c} = (\gamma_1,\ldots,\gamma_{n_k})^\top$, 
where $n_k = \lvert \SC^k \rvert$ is the number of $k$-simplices in the SC~(\cref{fig:schematic}).
Note that $\mathcal C_k$ is isomorphic to $\mathbb{R}^{n_1}$, so we may think of a chain as a vector in $\mathbb{R}^{n_1}$.

\setcounter{example}{0}
\begin{example}[continued]
    In \cref{fig:schematic_A}, the blue vector is the representation of the 1-chain $\bm{c} = (0,1,0,0,1,0,-2,0,-2,0)^\top$.
\end{example}

We make one further provision for the construction of $\mathcal C_k$---a change of the orientation of the basis element $s_i^k$ is defined to correspond to a change in the sign of the coefficient $\gamma_i$. 
Hence, if we ``flip'' a basis element $s^k_i$ to its opposite orientation, 
we have to multiply the corresponding coefficient $\gamma_i$ by $-1$.
Finally, we endow each space $\mathcal C_k$ with the standard $\ell^2$ inner product 
$\langle \bm{c}_1,\bm{c}_2\rangle = \bm{c}_1^\top\bm{c}_2$, and thus give $\mathcal C_k$
the structure of a finite-dimensional Hilbert space.

An alternative interpretation of the above construction is in terms of the space $\mathcal C^k$ of co-chains, which is the linear space of all alternating functions $f\colon\mathcal C_k \rightarrow \mathbb{R}$ (for a more detailed discussion, see Lim~\cite{Lim-2015-hodge}).\footnote{Strictly speaking, the space $\mathcal C^{k}$ of co-chains corresponds to the dual space of $\mathcal C_k$.}
Since $\mathcal C_k$ and $\mathcal C^k$ have the same dimension, and there is a canonical isomorphism between the space of chains $\mathcal C_k$ and co-chains $\mathcal C^k$, we will treat these two interpretations interchangeably in what follows even though their interpretation can be different.

The reader not familiar with these constructions may simply consider the above spaces in terms of their vector representation.
For instance, the space $\mathcal C^1$ can be interpreted as the space of \emph{edge-flows}, which are commonly encountered in graph theory.
Any vector $\bm f$ representing such a flow assigns one scalar value to each edge in the graph, where a negative value indicates that the flow is in the opposite direction with respect to the chosen reference orientation of the edge.
To illustrate the above mentioned duality of chains and co-chains for the edge-space, think of electrical circuits with unit resistances.
In this context, we may think of $\mathcal C^1$ as the space of edge-currents, and $\mathcal C_1$ as the space of edge-voltages, which encode exactly the same information.

\setcounter{example}{0}
\begin{example}[continued]
    The 1-chain $\bm{c} = (0,1,0,0,1,0,-2,0,-2,0)^\top$ in \cref{fig:schematic_A} 
    can by duality also be thought of as co-chain $\bm{f}$, or simply as the union of an edge-flow on 
    $2\rightarrow 6\rightarrow 5\rightarrow 4$ and an edge flow on $1\rightarrow 3$. 
    Due to the edge orientations, some entries of $\bm{c}$ are negative, corresponding to a flow in opposite direction of the reference orientation.
\end{example}

Another alternative way to think about the space of alternating functions on edges $\mathcal C^1$ 
is to identify it with the set of anti-symmetric matrices ($\bm{A} = -\bm{A}^\top$), 
whose sparsity pattern is consistent with the edges present ($A_{i,j} = A_{j,i} = 0$, if $\{i,j\} \notin \mathcal X^2$).
As $A_{i,j} = -A_{j,i}$, this representation simultaneously encodes both edge orientation $[i,j]$ and $[j,i]$ 
(with opposite signs, as desired).
In the following, we use the more compact description in terms of vectors, but the reader might find it insightful to keep the above picture in mind.

\begin{figure}[tb]
    \centering
    \begin{tikzpicture}[scale=0.9]
        \small
        \draw [fill=gray, opacity=0.3] (0,0) -- (3,0) -- (1.5,2) -- cycle;
        \tikzstyle{point}=[circle,thick,draw=black,fill=white,inner sep=2pt,minimum width=10pt,minimum height=10pt]
        \tikzstyle{smallpoint}=[circle,thick,draw=black,fill=black,inner sep=0pt,minimum width=5pt,minimum height=5pt]
        \begin{scope}[>=latex, thick, decoration={ markings, mark=at position 0.5 with {\arrow{>}}}]
            \node (a) at (0,0) [point] {1};
            \node (b) at (3,0) [point] {2};
            \node (c) at (1.5,2) [point] {3};
            \node (arrow) at (1.5,0.75) {$[1,2,3]$};

            \draw[postaction={decorate}] (a) -- node[below]{$[1,2]$} (b);
            \draw[postaction={decorate}] (b) -- node[right]{$[2,3]$} (c);
            \draw[postaction={decorate}] (a) -- node[left] {$[1,3]$} (c);
            \draw[thin,->] (2.15 ,0.75) arc (0:280:0.65cm);
        \end{scope}
        \begin{scope}[xshift=3cm,arrows=->,>=latex, thick]
            \node (start) at (0.25,0.75) {};
            \node (end) at (1.75,0.75) {};
            \draw (start) edge node[above]{$\partial_2$} (end);
        \end{scope}
        \begin{scope}[xshift=5cm,>=latex, thick, decoration={ markings, mark=at position 0.5 with {\arrow{>}}}]
            \node (a) at (0,0.3) [smallpoint] {};
            \node (a2) at (0.2,0) [smallpoint] {};
            \node (b) at (2.7,0) [smallpoint] {};
            \node (b2) at (3,0.3) [smallpoint] {};
            \node (c) at (1.3,2) [smallpoint] {};
            \node (c2) at (1.7,2) [smallpoint] {};

            \draw[postaction={decorate}] (a2) -- node[below]{$[1,2]$} (b);
            \draw[postaction={decorate}] (b2) -- node[right]{$[2,3]$} (c2);
            \draw[postaction={decorate}] (a) -- node[left] {$[1,3]$} (c);
        \end{scope}
        \begin{scope}[yshift=-1cm]
            \node (s3) at (1.5,0) {$[1,2,3]$};
            \node (start) at (3.25,0) {};
            \node (end) at (4.75,0) {};
            \draw (start) edge[->,>=latex,thick] node[above]{$\partial_2$} (end);
            \node (result) at (5,0) [right] {$[2,3] - [1,3] + [1,2]$};
        \end{scope} 
    \end{tikzpicture}
    \caption{\textbf{Illustration of the action of a boundary operator on a $2$-simplex.} 
    The boundary operator maps the $2$-simplex to a linear combination of its faces, respecting the orientation. 
    For simplicity, we choose a basis for $\mathcal C_k$ such that $\{s^k = [i_0,\ldots,i_k] : i_0 < \ldots < i_k\}$.
    Note that within the space of chains (see text), $-[1,3] = [3,1]$, which shows that the above boundary operator gives rise to a cycle $\partial_2([1,2,3]) = [1,2] + [2,3] + [3,1]$ for which $\partial_1(\partial_2([1,2,3])) = 0$.
    }\label{fig:schematic1}
\end{figure}
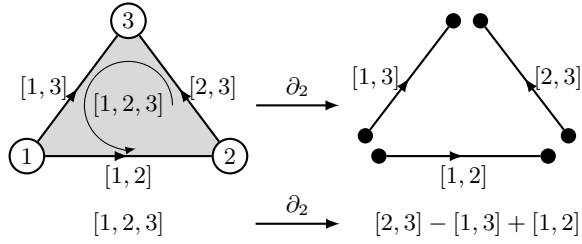

\subsection{Boundary and co-boundary maps}
Given the spaces of chains $\mathcal C_k$ defined above, we define the linear boundary maps $\partial_k\colon~\mathcal C_k~\rightarrow~\mathcal C_{k-1}$ by their action on the basis elements as follows:
\begin{equation*}
    \partial_k([i_0,\ldots,i_k]) = \sum_{j=0}^k (-1)^j[i_0,\ldots,i_{j-1},i_{j+1},\ldots,i_k].
\end{equation*}
These operators map any chain to a sum of its boundary components, i.e., the simplices lower adjacent to the $k$-chain considered, with the appropriate orientation (\cref{fig:schematic1}).
We thus call $\text{im}(\partial_k)$ the space of $(k-1)$-boundaries, where $\text{im}(\cdot)$ denotes the image of an operator.
It is not difficult to show that if we build a cyclic chain $c_k\in\mathcal C_k$ 
whose start- and end-points are identical, then $\partial_k c_k = 0$; 
and similarly, any $c_k$ such that $\partial_k c_k = 0$ must correspond to a cycle.
Thus, we call $\text{ker}(\partial_k)$ the space of $k$-cycles.

The boundary operators are linear maps between finite-dimensional vector spaces. 
After choosing a basis, each of these operators can be represented by a matrix, thereby enabling 
us to perform computations based on these objects.
We will denote the matrix representation of the boundary operators 
$\partial_k$ by $\bm{B}_k$ 
(see \cite{Muhammad2006,Tahbaz-Salehi2010,Lim-2015-hodge} for further discussion on how to construct these matrices).

\setcounter{example}{0}
\begin{example}[continued]
    Consider again the simplicial complex in \cref{fig:schematic}. In this case, the boundary maps $\bm{B}_1$ (rows indexed by nodes, columns indexed by edges) and $\bm{B}_2$ (rows indexed by edges, columns indexed by $2$-simplices) are:
    \begin{equation*}
        \small
        \arraycolsep=1.4pt\def\arraystretch{1}
        \bm{B}_1 = 
        \begin{array}{c|cccccccccc}
            \;& [1,2] & [1,3] & [2,3] & [2,4] & [2,6] & [3,4] & [4,5] & [4,7] & [5,6] & [5,7]\\
                \hline 
             1  &   -1 & -1 & 0 & 0 & 0 & 0 & 0 & 0 & 0 & 0\\
             2  &   1 & 0 & -1 & -1 & -1 & 0 & 0 & 0 & 0 & 0\\
             3  &   0 & 1 & 1 & 0 & 0 & -1 & 0 & 0 & 0 & 0\\
             4  &   0 & 0 & 0 & 1 & 0 & 1 & -1 & -1 & 0 & 0\\
             5  &   0 & 0 & 0 & 0 & 0 & 0 & 1 & 0 & -1 & -1\\
             6  &   0 & 0 & 0 & 0 & 1 & 0 & 0 & 0 & 1 & 0\\
             7  &   0 & 0 & 0 & 0 & 0 & 0 & 0 & 1 & 0 & 1\\
            \end{array}\;\;\;
        \bm{B}_2 = 
        \begin{array}{c|cc}
            \; & [1,2,3] & [2,3,4] \\
            \hline
            [1, 2] & 1 & 0\\
            \left [1, 3\right] & -1 & 0\\
            \left [2,3\right ] & 1 & 1\\
            \left [2,4\right ] & 0 & -1\\
            \left [2,6\right ] & 0 & 0\\
            \left [3,4\right ] & 0 & 1\\
            \left [4,5\right ] & 0 & 0\\
            \left [4,7\right ] & 0 & 0\\
            \left [5,6\right ] & 0 & 0\\
            \left [5,7\right ] & 0 & 0\\
        \end{array}
    \end{equation*}
    The matrix $\bm{B}_1$ is the node-to-edge incidence matrix from algebraic graph theory. 
    Likewise, the higher-order boundary maps induce matrices $\bm{B}_i$ that can be interpreted 
    as higher-order incidence matrices between simplices and their (co-)faces.
\end{example}

Note that for each boundary map there exists a co-boundary map $\partial_k^\top\colon \mathcal C_k \rightarrow \mathcal C_{k+1}$, which is simply the adjoint of the boundary map.
The matrix representation of the co-boundary operator $\partial_k^\top$ is $\bm{B}_k^\top$.

\subsection{Hodge Laplacians}\label{sec:higher_order_laplacians}
From the sequences of boundary maps, one can define a hierarchy of Laplacian operators for the SC $\SC$.
Using our matrix representations discussed above, the $k$th \emph{combinatorial Hodge Laplacian} is:
\begin{equation}
    \bm{L}_k = \bm{B}_k^\top\bm{B}_k^{} + \bm{B}_{k+1}^{}\bm{B}_{k+1}^\top.
\end{equation}
The standard combinatorial graph Laplacian is a special case of the above and corresponds to 
$\bm{L}_0 = \bm{B}_1 \bm{B}_1^\top$ (as $\bm{B}_0 =0$).
The matrix $\bm{L}_1$, which is also referred to simply as the (Hodge) $1$-Laplacian, 
is the primary focus of this paper.

As solutions to the Laplace equation ($\Delta\bm{x} = 0$) are called harmonic functions, and $\bm{L}_k$ 
may be interpreted as a discretized version of the Laplace equation~\cite{Lim-2015-hodge}, 
elements $\bm{h} \in \text{ker}(\bm L_k)$ are called harmonic (functions).
These harmonic functions also carry a specific topological meaning.
From the definitions of the boundary maps, we can compute that $\partial_{k-1}\circ\partial_{k} = 0$.
Thus, the adjoint of this map is also zero, i.e., $\partial_{k}^\top\circ\partial_{k-1}^\top =0$.
These equations encapsulate the natural idea that the ``boundary of a boundary'' is empty.
In matrix terms, this annihilation of the boundary maps translates into 
$\bm{B}_{k}\bm{B}_{k+1} = 0$ and $\bm{B}_{i+1}^\top\bm{B}_{i}^\top = 0$.

Furthermore, since $\partial_{k}\circ \partial_{k+1} = 0$, $\text{im} (\partial_{k+1})$ is a subspace of $\text{ker}(\partial_{k})$.
This leads to the definition of the homology vector spaces of $\SC$ over $\mathbb{R}$, as those elements in the null space $\text{ker}(\partial_k)$ which are not in the image $\text{im} (\partial_{k+1})$:
\begin{equation}
    \mathcal H_k := \mathcal H(\SC,\mathbb{R}) = \text{ker}(\partial_k) \big/ \text{im}(\partial_{k+1}).
\end{equation}
Intuitively, the homology $\mathcal H_k$ may be interpreted as accounting for the number of $k$-dimensional ``holes'' in the SC $\SC$.
More precisely, elements of $\mathcal H_k$ correspond to $k$-cycles that are not induced by a $k$-boundary.
The number of $k$-dimensional holes in the simplicial complex is the so called the $k$th Betti number.
It can be shown that this corresponds precisely to the dimension of the null space of the $k$th Hodge Laplacian $\text{ker} (\bm L_k)$~\cite{Lim-2015-hodge}.

\subsection{The Hodge decomposition}
The combinatorial Hodge Laplacian is a sum of two positive semi-definite
operators, so any $\bm{h} \in \text{ker}(\bm{L}_k)$ fulfills
$\bm{h} \in \text{ker}(\bm{B}_k)$ and
$\bm{h} \in \text{ker}(\bm{B}_{k+1}^\top)$.
This implies that the nonzero elements in $\text{ker}(\bm L_k)$
are representatives of the non-trivial equivalence classes in the $k$th homology.

The space $\mathcal C_k$ is isomorphic to $ \mathbb{R}^{n_k}$ within our chosen representation,
which can be represented by
$\text{ker}(\bm L_k) \oplus \text{im}(\bm L_k^T) = \text{ker}(\bm L_k) \oplus \text{im}(\bm L_k)$,
where $\oplus$ denotes the union of orthogonal subspaces with respect to the standard inner product.
Clearly, $\text{im}(\bm L_k) \subseteq \text{im}(\bm B_k^T) \cup \text{im}(\bm B_{k+1})$,
Moreover, as $\text{im}(\bm B_{k + 1}) \subset \text{ker}(\bm B_{k}^T)$ in our case,
we have that $\text{im}(\bm L_k) \subseteq \text{im}(\bm B_k^T) \oplus \text{im}(\bm B_{k+1})$.
Since the dimension of $\text{im}(\bm B_k^T) \oplus \text{im}(\bm B_{k+1})$ cannot exceed the dimension
of $\text{im}(\bm L_k)$, these subspaces must be equal, and we can decompose $C_k$ as follows:
\begin{equation}\label{eq:Hodge}
  \mathcal C_k \simeq \mathbb{R}^{n_k} =
  \text{im}(\bm{B}_{k+1}) \oplus \text{im} (\bm{B}_k^\top)  \oplus \text{ker}(\bm L_k).
\end{equation}
\Cref{eq:Hodge} is called the \emph{Hodge decomposition}.
Later, we discuss how the Hodge decomposition for $\mathcal C_1$ can provide additional insights into data.
 
\section{Diffusion processes on simplicial complexes}\label{sec:diffusion_on_SC}
In this section, we outline our model for diffusion processes on SCs that accounts for topological features.
For simplicity, we focus on $1$-simplices, i.e., diffusion between edges.

\paragraph{Diffusion processes on graphs}
To understand the complications of defining a diffusion process on an SC, let us revisit a random walk on a graph, a prototypical model for a diffusion process on a graph.
A (standard, unbiased) random walk on a graph with adjacency matrix $\bm{A}$ can be described by the following transition rule:
\begin{equation}\label{eq:random_walk_graph}
    \bm{p}_{t+1} = \bm{AD}^{-1}\bm{p}_t = (\bm{I} - \bm{L}_0\bm{D}^{-1})\bm{p}_t.
\end{equation}
Here the $i$th component of the vector $\bm{p}_t$ denotes the probability of finding a random walker at node $i$ at time $t$, and $\bm{p}_0$ corresponds to an initial distribution of the walker.

There are two important features of this formulation.
First, the transition matrix of the random walk is directly related to a normalized Hodge Laplacian, namely, 
$\bm{\mathcal{L}}_0 = \bm{L}_0\bm{D}^{-1}$ is the so-called random walk Laplacian.
There is thus a close relationship between the topological features of the graph and the random walk,
as the harmonic functions of $\bm{\mathcal{L}}_0$ are directly related to the connected components of the graph.
Second, the state space and the transitions of the random walker is determined by the graph.
The nodes are the states of the random walker and transitions occur over the edges.

\subsection{Beyond graphs: keeping track of orientations}
When extending the  concept of a random walk to SCs, 
a mismatch between the two features discussed above becomes apparent if we go beyond the node-space.
On the one hand, we may define a random walk on the edges, 
where the edges themselves are defined as the states of the Markov process.
To define such a process we could use the line graph~\cite{Evans2009,Ahn2010}
or other ``dual graph'' constructions~\cite{Osting2017}.
However, we would abandon the connection to algebraic topology and the Hodge Laplacian, 
as the Laplacian of the line graph is not directly related to the Hodge Laplacian of the SC.
The properties of a random walk on the line graph will therefore not be informative about the topology of the SC.

On the other hand, we face a different issues if we define a random walk based on the 
$\bm{L}_1$ Laplacian formally analogous to \cref{eq:random_walk_graph}.
The Laplacian $\bm{L}_1$ has non-trivial patterns of positive and negative entries that depend on the edge orientations.
Hence, the $\bm{L}_1$ Laplacian is not obviously related to a transition matrix of a Markov chain.
We do not face this issue with the $\bm{L}_0$ Laplacian because orientation is trivial for vertices.

Can a ``normalized'' variant of the $\bm{L}_1$ Laplacian be related to a random walk?
It turns out that we can indeed construct an edge-based diffusion process that is tied to the topology of the original complex.
However, we have to consider a random walk in a higher-dimensional, lifted state space.
Our idea is that instead of considering how $\bm{L}_1$ acts on an edge-flow $\bm{f}$ in the original space, 
we view its action as an equivalent sequence of three operations:
first, we lift $\bm{f}$ into a higher dimensional space; second, we act on it via a linear operator; and third, we project the result back down to the original state space.
Once we understand these actions, we can normalize the linear transformation in the lifted space such that it corresponds to a diffusion.
This leads to the definition of a normalized Hodge Laplacian, to which we can assign a meaning in terms of a random walk in a lifted space.

Decomposing the action of the Hodge Laplacian $\bm L_1$ in the above way (lift, apply, project) 
enables us to disentangle the orientation of a flow with the magnitude of the flow.
The magnitude of each component indicates the volume of the flow; whereas the sign of the variable indicates the direction of the flow, which can be aligned or anti-aligned with our chosen reference orientation.
As we will see, the magnitude of the flow that can be related to a probability, whereas the information about the direction of the flow is a matter of accounting for a reference orientation.

\subsection{Lifting of edge-flows and matrix operators}\label{ss:liftings}
\begin{figure}[tb!]
  \phantomsubfigure{fig:lifting_A}
  \phantomsubfigure{fig:lifting_B}
   \phantomsubfigure{fig:lifting_C}
    \centering
    \includegraphics[]{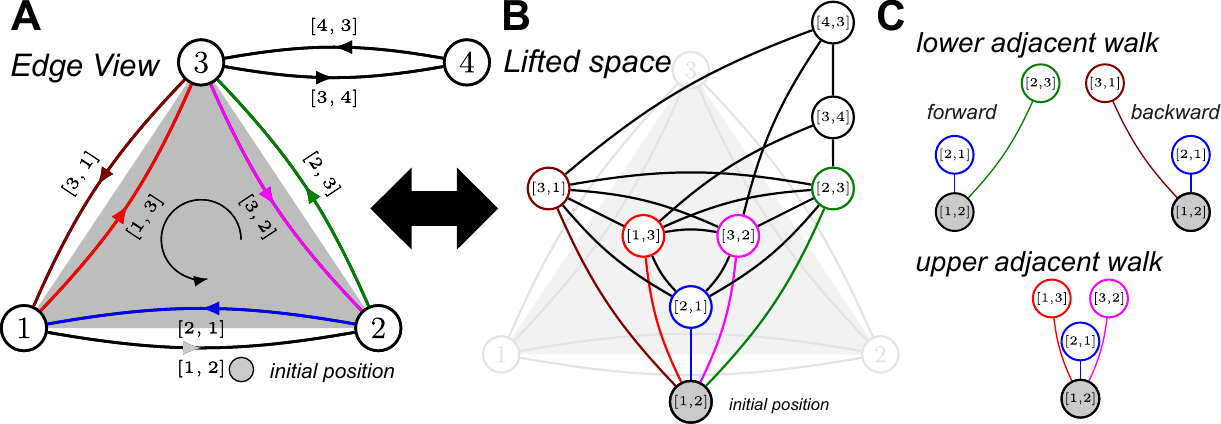}
    \caption{\textbf{Illustration of a lifted simplicial complex.} 
    \textbf{A} We may think of the of lifted complex as an augmented complex in which each original edge is represented in both possible orientations. \textbf{B}~Alternatively, we may interpret each oriented edge $[i,j]$ in the original complex, as giving rise to two states $[i,j]$ and $[j,i]$ on a graph with $2n_1$ vertices. 
    \textbf{C} Starting from $[1,2]$, there are lower adjacent connections ``forward'' and 
    ``backward,'' as well as upper adjacent connections (see text).}
    \label{fig:lifting}
\end{figure}

In the following, we describe how the action of the $\bm{L}_1$ Laplacian on any co-chain vector $\bm{f}$ (edge-flow) can be understood from the point of view of a higher-dimensional, lifted state space.
We consider a lifting of an edge-flow $\bm{f} \in \mathcal C^1$ into a larger space $\mathcal D^1$ in which both possible orientations for each edge are present (\cref{fig:lifting}).
Since there are two possible orientations for each edge, $\lvert \mathcal D^1\rvert = 2 \lvert \mathcal C^1\rvert$.
As an edge-flow in $\mathcal C^1$ corresponds to an alternating function, there is a natural inclusion map $V\colon \mathcal C^1 \rightarrow \mathcal D^1$ which maps any edge-flow into $\mathcal D^1$ by explicitly representing both edge directions.
We choose the basis elements of $\mathcal D_1$ such that the matrix representation of $V$ is: 
\begin{equation}
    \bm{V} = \begin{pmatrix} +\bm{I}_{n_1} \\ -\bm{I}_{n_1} \end{pmatrix} \in \mathbb{R}^{2n_1 \times n_1},
\end{equation}
where $\bm{I}_{n_1}$ is the identity matrix of dimension $n_1 = \lvert \mathcal C_1 \rvert$.

\setcounter{example}{0}
\begin{example}[continued]
    Consider the edge-flow $\bm{f} = (0,1,0,0,1,0,-2,0,-2,0)^\top$ in \cref{fig:schematic_A}. 
    The lifted edge-flow is simply $\bm{\widehat{f}} = \bm{Vf} = (\bm{f}^\top, -\bm{f}^\top)^\top$. 
    For instance, $\bm{\widehat{f}}$ has an entry $1$ for edge $[2, 6]$ and $-1$ for the (now added) reverse edge $[6, 2]$.
\end{example}

The lifting operator has the property that $\bm{V}^\top \bm{V} = 2 \bm{I}_{n_1}$.
Thus, the Moore--Penrose pseudoinverse of $\bm{V}$ is $\bm{V}^\dagger =\frac{1}{2} \bm{V}^\top$.
Furthermore, it is easy to see that $\bm{V}\bm{V}^\top = \bm{I}_{2n_1} - \bm{\Sigma}$, where $\bm{\Sigma}$ is the permutation matrix that maps the original basis simplices to their counterparts with switched orientation:
\begin{equation}
    \bm{\Sigma} = \begin{pmatrix} \bm{0} & \bm{I}_{n_1}\\ \bm{I}_{n_1} & \bm{0} \end{pmatrix}.
\end{equation}

Having defined a lifting for an edge-flow, we now need to define an appropriate notion for a lifting of a matrix operator.
\begin{definition}[Lifting of a matrix]
    We say that a matrix $\bm{N} \in \mathbb{R}^{2n_k \times 2n_k}$ is a lifting of a 
    matrix $\bm{M} \in \mathbb{R}^{n_k \times n_k}$ if the following condition holds:
    \begin{equation}
         \bm{V}^\top \bm{N} = \bm{M}\bm{V}^\top.
    \end{equation}
\end{definition}
This definition implies that if a matrix $\bm{M}$ has a lifting $\bm{N}$,
then by multiplying from the right with $\bm{V}$, $\bm{M} = \frac{1}{2}\bm{V}^\top\bm{N}\bm{V}= \bm{V}^\dagger \bm{N}\bm{V}$.
Hence, the action of multiplying by $\bm{M}$ can be interpreted in terms of a lifting, followed by a linear
transformation, and finally a projection into the original lower-dimensional
space (we use the term ``projection'' here to refer to a mapping into a lower-dimensional space).

We now consider a lifting of the combinatorial Hodge Laplacian $\bm L_1$.
To state our results compactly we define the following matrices :
\begin{equation}
    \liftB_1 := \bm{B}_1\bm{V}^\top = \begin{pmatrix}\bm{B}_1 &  -\bm{B}_1 \end{pmatrix}
    \quad\text{and}\quad 
    \liftB_2 := \bm{V}\bm{B}_2 = 
    \begin{pmatrix}
    \bm B_2\\ - \bm B_2
    \end{pmatrix}.
\end{equation}
We will moreover make use of the positive part $\liftB_i^+$ and the negative part $\liftB_i^-$ of these matrices.
In the following we use the notation $\bm{\widehat{\cdot}}$ to indicate objects that are related to the lifted space.
Note that such objects may not always be liftings of matrices~(consider, e.g., $\liftB_1$).

\begin{lemma}[Lifting of the $\bm{L}_1$ Hodge Laplacian]\label{lem:lifting1}
    The negative of the Hodge Laplacian $\bm{L}_1 = \bm{B}_1^\top\bm{B}_1 + \bm{B}_2\bm{B}_2^\top$ 
    has a lifting $\bm{\widehat{A}} = \bm{\widehat{A}}_l +  \bm{\widehat{A}}_u$ with:
    \begin{equation*}
        \bm{\widehat{A}}_l= (\liftB_1^-)^\top \liftB_1^+ + (\liftB_1^+)^\top \liftB_1^- \text{ and }  \widehat{\bm{A}}_u = \liftB_2^+(\liftB_2^-)^\top  + \liftB_2^-(\liftB_2^+)^\top.
    \end{equation*}
\end{lemma}

Let us unpack this result before proving it.
We can immediately see that $\bm{\widehat{A}}$ is symmetric.
Next, for any matrix $\bm{M}$, we have that $(-\bm{M})^{-} = \bm{M}^+$; $(-\bm{M})^{+} = \bm{M}^{-}$; and $\bm{M}^{+}$ and $\bm{M}^{-}$ are nonnegative.
Thus, $\bm{\widehat{A}}$ is also non-negative and can be interpreted as the (weighted) adjacency matrix of an undirected graph with $2n_1$ nodes.
More specifically, $\bm{\widehat{A}}_l$ describes connections between lower adjacent edges and is composed of:
\begin{equation*}
    ((\liftB_1^-)^\top \liftB_1^+)_{[i,j],[k,l]} = \begin{cases} 1 & \text{if } l=i \\ 0 & \text{otherwise}\end{cases} \quad
    ((\liftB_1^+)^\top \liftB_1^-)_{[i,j],[k,l]} = \begin{cases} 1 & \text{if } k=j \\ 0 & \text{otherwise.}\end{cases} 
\end{equation*}
The first matrix describes a forward walk respecting the edge orientation, 
where the target node $l$ of the first edge $[k,l]$ has to match with source-node $i$ of the second edge $[i,j]$.
The second matrix describes a backward walk in the opposite direction, 
where the source node $k$ of the first edge $[k,l]$ has to match with the target-node $j$ of the second edge $[i,j]$ (\cref{fig:lifting}).

Likewise, $\widehat{\bm{A}}_u$ describes connections between upper adjacent edges with a joint triangular co-face.
Using the symbol $\nsim$ to denote that two edges have a different orientation relative to a joint co-face, we can write:
\begin{align}
    (\widehat{\bm{A}}_u)_{[i,j],[k,l]} = \begin{cases} 1 & \text{if } [k,l] \nsim [i,j] \\ 0 & \text{otherwise.}\end{cases}
    \label{eq:A_upper}
\end{align}

\begin{example}
    Consider \Cref{fig:lifting}. In the lifted space $\mathcal D^1$, the lower adjacent forward connections of 
    $[1,2]$ are $[2,1]$ and $[2,3]$; the lower adjacent backward connections are $[2,1]$ and $[3,1]$.
    The upper adjacent connections of $[1,2]$ with opposite orientation are $[2,1]$, $[3,2]$, and $[1,3]$.
\end{example}
To conclude this section, we prove \cref{lem:lifting1}
\begin{proof}
    Since $(-\bm{M})^- = \bm{M}^+$ and $(-\bm{M})^+ = \bm{M}^-$, for $\widehat{\bm{A}}_l$ we have that:
    \begin{align*}
        \liftB_1^-\bm{V} = (\bm{B}_1, -\bm{B}_1)^-\;\bm{V} = -\bm{B}_1, \quad \liftB_1^+\bm{V} = (\bm{B}_1, -\bm{B}_1)^+\;\bm{V} = \bm{B}_1, \quad
        \liftB_1^+ \bm\Sigma &= \liftB_1^-.
    \end{align*}
    
    Using the transposes of the first two equalities as well as the third equality,
    \begin{align*}
        \bm{V}^\top \widehat{\bm{A}}_l &= - \bm{B}_1^\top \widehat{\bm{B}}_1^+ + \bm{B}_1^\top \widehat{\bm{B}}_1^- = -\bm{B}_1^\top\widehat{\bm{B}}_1^+(\bm{I}- \bm\Sigma)
        = -\bm{B}_1^\top\widehat{\bm{B}}_1^+\bm{V}\bm{V}^\top= -\bm{B}_1^\top\bm{B}_1\bm{V}^\top.
    \end{align*}

    By analogous arguments for $\widehat{\bm{A}}_u$, we obtain:
    \begin{align*}
        \bm{V}^\top \widehat{\bm{A}}_u &= \bm{B}_2(\widehat{\bm{B}}_2^-)^\top - \bm{B}_2(\widehat{\bm{B}}_2^+)^\top
                                                   = - \bm{B}_2(\widehat{\bm{B}}_2^+)^\top(\bm{I} - \bm\Sigma) = - \bm{B}_2\bm{B}_2^\top\bm{V}^\top.
    \end{align*}
    The lemma follows by combining these two results.
\end{proof}

\subsection{The normalized Hodge 1-Laplacian and edge-space random walks}
Motivated by our lifting result in~\Cref{lem:lifting1}, we now define a normalized Hodge Laplacian for the edge-space and show that its action can be related to a random walk on a lifted complex.
There is some flexibility here, however, as multiple operators in the lifted space $\mathcal D^1$ will correspond to the same projected matrix.
Likewise, there are multiple types of random walks we could define in the lifted space by assigning different weights to the various transitions, leading to different notions of a normalized Laplacian operator.
The normalized Hodge Laplacian we consider here is of a ``standard form''~\cite{Horak2013,grady2010discrete} and admits a (normalized) Hodge decomposition, which is of interest for our applications.
A systematic exploration of further normalization schemes and their respective 
advantages is an interesting avenue for future research.

\begin{definition}[normalized Hodge 1-Laplacian $\bm{\mathcal{L}}_1$]
    Consider a simplicial complex $\SC$ whose boundary operators can be represented by the matrices $\bm B_1$ and $\bm B_2$.
    The normalized Hodge 1-Laplacian matrix is then defined by
    \begin{equation}\label{eq:def_hodge_laplacian}
        \bm{\mathcal{L}}_1 = \bm{D}_2\bm{B}_1^\top\bm{D}_1^{-1}\bm{B}_1^{} + \bm{B}_2^{}\bm{D}_3\bm{B}_2^\top \bm{D}_2^{-1},
    \end{equation}
    where $\bm{D}_\mathrm{2}$ is the diagonal matrix of (adjusted) degrees of each edge:
    \begin{equation}
	\bm{D}_\mathrm{2} = \max(\textnormal{diag}(\lvert \bm{B}_2 \rvert \bm{1}), \bm{I}) \iff
	(\bm{D}_\mathrm{2})_{[i,j],[i,j]} = \max\{\mathrm{deg}([i,j]), 1\},
    \end{equation}
    $\bm{D}_1 = 2 \cdot \textnormal{diag} (\lvert \bm{B}_1 \rvert \bm{D}_2 \bm{1})$
    is a diagonal matrix of weighted degrees of the nodes
    (with the weight of an edge equal to the maximum of $1$ and the number of co-faces of the edge),
    and $\bm{D}_3 = \frac{1}{3}\bm{I}$.
\end{definition}

In the above definition, the matrix $\bm{D}_2$ defines a weighting of the edges according to their degree, 
where the element-wise maximum in $\bm{D}_2$ ensures that the normalized Hodge Laplacian is well-defined 
(i.e., the edge weight of an existing edge is at least 1).
The matrix $\bm{D}_1$ encodes twice the weighted degree of the nodes according 
to the weights of the incident edges, and
$\bm{D}_3$ gives a weighting of $1/3$ to each triangular face.
Recall that in the standard case of graphs, the Hodge Laplacian is
\[
\bm{L}_0 = \bm{0}_1^\top\bm{0}_1^{} + \bm{B}_1^{}\bm{B}_1^\top = \bm{B}_1^{}\bm{B}_1^\top,
\]
and the normalized (random walk) version is
\[
\bm{\mathcal{L}}_0 =  \bm{B}_1^{}\widetilde{\bm{D}}_3\bm{B}_1^\top\widetilde{\bm{D}}_2^{-1},
\]
with $\widetilde{\bm{D}}_3 = \bm{I}$ and $\widetilde{\bm{D}}_2 = \max(\textnormal{diag}( \lvert \bm{B_1} \rvert \bm{1}), \bm{I})$.
Note how $\widetilde{\bm{D}}_2$ and $\bm{D}_2$ have exactly the same functional form, but are based on $\bm B_1$ and $\bm B_2$, respectively.
Similarly, both $\widetilde{\bm{D}}_3$ and $\bm{D}_3$ are simple scalings.

The key difference between the normalized random walk Laplacian and the first-order normalized Hodge Laplacian is that the latter contains both an upper adjacent as well as a lower adjacent term.
The following result shows that this normalized Hodge 1-Laplacian has a meaningful connection to random walks.
Specifically, the matrix is related to a random walk in the lifted space of edges.
\begin{theorem}[Stochastic Lifting of the normalized Hodge 1-Laplacian]\label{thm:stoch_lifting}
    The matrix $-\bm{\mathcal{L}}_1/2$ has a stochastic lifting, i.e., there exists a column stochastic matrix 
    $\bm{\widehat P}$ such that $-\frac{1}{2}\bm{\mathcal{L}}_1\bm{V}^\top = \bm{V}^\top \bm{\widehat P}$.
    Specifically, $\bm{\widehat P} := \frac{1}{2}\bm{P}_\textnormal{lower} + \frac{1}{2}\bm{P}_\textnormal{upper}$,
    where $\bm{P}_\textnormal{lower}$ is the transition matrix of a random walk determined by the lower-adjacent connections and 
    $\bm{P}_\textnormal{upper}$ is the transition matrix of a random walk determined by the upper-adjacent connections.
    The transition matrix $\bm{P}_\textnormal{lower}$ is defined by a ``forward walk'' and ``backward walk'' component moving in the orientation of the edges or against it, respectively:
    \begin{align}
        \bm{P}_\textnormal{lower} := \frac{1}{2}\left( \bm{P}_\textnormal{lower,forward} + \bm{P}_\textnormal{lower,backward} \right) \\ 
        \bm{P}_\textnormal{lower,forward} =  \bm M_f \textnormal{diag}(\bm{M}_f \bm{1})^{-1}\\
        \bm{P}_\textnormal{lower,backward} = \bm M_b \textnormal{diag}(\bm{M}_b \bm{1})^{-1}
    \end{align}
    where $\bm M_f = \bm{\widehat{D}}_2(\liftB_1^-)^\top \liftB_1^+$ and $\bm M_b = \bm{\widehat{D}}_2(\liftB_1^+)^\top \liftB_1^-$ are (weighted) lower adjacency matrices corresponding to forward and backward walks along the edges (see~\Cref{lem:lifting1})
    and $\bm{\widehat{D}}_2 = \textnormal{diag}(\bm{D}_\mathrm{2},\bm{D}_\mathrm{2})$.
    The transition matrix $\bm{P}_\textnormal{upper}$ describes a random walk along upper adjacent faces as follows:
    \begin{align}
        \bm{P}_\textnormal{upper} = \widehat{\bm{A}}_u \widehat{\bm{D}}_4^{-1}
        + \frac{1}{2} \begin{pmatrix} \bm I & \bm  I \\ \bm I & \bm I \end{pmatrix} \widehat{\bm D}_5,
    \end{align}
    where $\widehat{\bm{A}}_u = \liftB_2^+(\liftB_2^-)^\top  + \liftB_2^-(\liftB_2^+)^\top$
    is the matrix of upper adjacent connections as defined in \cref{lem:lifting1}, and $\widehat{\bm{D}}_4$ is a diagonal matrix:
    \begin{equation}
      (\widehat{\bm{D}}_4)_{[i,j],[i,j]} =
      \begin{cases}
        1 & \text{if} \quad \mathrm{deg}([i,j]) = 0\\ 
        3 \cdot \mathrm{deg}([i,j]) & \text{otherwise}.
      \end{cases}
    \end{equation}
    Here $\widehat{\bm D}_5$ is the diagonal matrix selecting all edges with no upper adjacent faces:
    \begin{equation}
      (\widehat{\bm{D}}_5)_{[i,j],[i,j]} =
      \begin{cases} 1 & \text{if} \quad \mathrm{deg}([i,j]) = 0
        \\ 0 & \text{otherwise}.
      \end{cases}
    \end{equation}
\end{theorem}

\begin{proof}
    The proof closely follows our lifting result above and is in the appendix.
\end{proof}

The random walk described by $\widehat{\bm P}$ can be described in words as follows. 
With a probability of $0.5$ each, we take a step via either the upper or lower adjacent connections. 
If we take a step via the lower adjacent connections (via the nodes), then with probability of $0.5$ each we move either along or against the chosen edge orientation.
In either case, the transition probability to a target edge is then proportional to the upper degree of the target edge, which corresponds to the ``weight'' of that edge.
If we take a step via the upper adjacent connection there are two cases.
If the edge has no upper adjacent face, then the random walk will stay at the same oriented edge or change orientation with equal probability $0.5$.
If the edge has an upper adjacent face, then a walker on edge $[i,j]$ will 
transition uniformly to an upper adjacent edge $[k,l] \nsim [i,j]$ with different orientation relative to their shared face.
Stated differently, the walker performs an unbiased random walk on the lifted graph with adjacency matrix $\widehat{\bm{A}}_u$, unless there is no upper adjacent connection in which case the walker will either stay put or move to the edge with reverse orientation.

Finally, there exists an interesting link between the construction of the normalized Hodge 1-Laplacian here and previous 
research that considered certain relationship between Hodge Laplacians and \emph{differences} of random walks on simplicial complexes 
(with absorbing states)~\cite{Mukherjee2016,Rosenthal2014}.
As the projection operator computes a difference between the two possible orientations of an edge in the lifted space, a question for future work would be to explore this connection in more detail.

\paragraph{Spectral properties}
The ideas underpinning \cref{thm:stoch_lifting} enable us to derive the following results, 
which have consequences for the spectral properties of $\bm{\mathcal{L}}_1$.
\begin{corollary}\label{thm:relations}
    Define the matrix $\bm{Z} = - \bm{\mathcal{L}}_1/2$ and the matrix $\widehat{\bm{P}}$ as in \cref{thm:stoch_lifting}.
    Then the following identities hold:
    \begin{enumerate}
    \item $\bm{ZV}^\top = \bm{V}^\top\bm{\widehat{P}}$,
    \item $\bm{Z} = \bm{V}^\dagger \widehat{\bm{P}}\bm{V}$, and
    \item $\bm{VZ} = \bm{VV}^\dagger\widehat{\bm{P}}\bm{V} = \bm{\widehat{P}}\bm{V}$.
    \end{enumerate}
\end{corollary}
\begin{proof} The first two relations are a simple restatement of \cref{thm:stoch_lifting}. 
The last equality can be shown analogously to \cref{thm:stoch_lifting} and is omitted for brevity.
\end{proof}

\begin{corollary}
    Consider the (lifted) space $\mathcal D_1$ of edge flows (cf.~\cref{ss:liftings}).
    The subspace of alternating functions $\text{span}(\bm{V}) \subset \mathcal D_1$ is an invariant subspace of $\widehat{\bm{P}}$.
    Consequently, the spectrum of $\bm{Z}$ is contained in the spectrum of $\bm{\widehat{P}}$, 
    i.e., $\lambda(\bm{Z}) \subset \lambda(\bm{\widehat{P}})$.
    Furthermore, any eigenvector $\bm{x}$ of $\bm{Z}$ with eigenvalue $\lambda$ 
    corresponds to an eigenvector of $\bm{\widehat{P}}$ of the form $\bm{y} = \bm{V}\bm{x}$ with the same eigenvalue.
\end{corollary}
\begin{proof}
    The first claim follows immediately from $\bm{\widehat{P}}\bm{V} =\bm{VZ}$.
    The latter parts follow by using the identities established in \cref{thm:relations} to compute:
    $\bm{\widehat{P}}\bm{y} = \bm{\widehat{P}}\bm{Vx}=\bm{V}\bm{Z}\bm{x} = \lambda\bm{V}\bm{x}$, which holds for any eigenvector $\bm{x}$ of $\bm{Z}$.
\end{proof}
The above results implies that, like the normalized graph Laplacian, the spectrum of $\bm{\mathcal{L}}_1$ has a bounded support.

\subsection{Normalized Hodge decompositions}\label{sec:norm_hodge_lap}
Similar to the Hodge 1-Laplacian, the eigenvectors of the normalized Hodge 1-Laplacian 
$\bm{\mathcal{L}}_1$ associated to the eigenvalue $\lambda=0$ of $\bm{\mathcal{L}}$ and 
the induced eigenvectors of $\bm{\widehat{P}}$ are associated with scaled harmonic functions. 
In fact, we can obtain the following normalized (weighted) Hodge decomposition from our normalized Hodge 1-Laplacian:
\begin{equation}\label{eq:hodge_skew}
    \mathbb{R}^{n_1} = \text{im}(\bm{B_2}) \oplus_{\bm D_2^{-1}} \text{im}(\bm{D}_2\bm{B}_1^\top) \oplus_{\bm D_2^{-1}} \text{ker}(\bm{\mathcal{L}}_1),
\end{equation}
where $\oplus_{\bm D_2^{-1}}$ denotes the union of orthogonal subspaces with respect to the inner product $\langle \bm x, \bm y \rangle_{\bm D_2^{-1}} = \bm x^\top \bm D_2^{-1} \bm y$.
Comparing \cref{eq:Hodge} with \cref{eq:hodge_skew}, it should be apparent that there is an isomorphism between the respective subspace of the standard and the normalized Hodge Laplacian and thus a correspondence between the harmonic functions associated with $\bm L_1$ and $\nLap_1$.
If we consider a symmetrized version of $\bm{\mathcal{L}}_1$ given by $\bm{\mathcal{L}}_1^\text{s} = \bm D_2^{-1/2}\bm{\mathcal{L}}_1\bm D_2^{1/2}$, the corresponding Hodge decomposition holds again with respect to the standard inner product:
\begin{equation}\label{eq:Hodge_weighted}
    \mathbb{R}^{n_1} = \text{im}(\bm D_2^{-1/2}\bm{B}_2) \oplus \text{im}(\bm{D}_2^{1/2}\bm{B}_1^\top)\oplus \text{ker}(\bm{\mathcal{L}}_1^\text{s}).
\end{equation}
We will use this normalized Hodge decomposition in our application examples 
(see also~\Cref{sec:hodge_background} for further discussion).

In addition to the eigenvectors associated to $0$ eigenvalues, which have a
clear interpretation in terms of harmonic functions (homology), we can provide
further insights on the remaining eigenvectors. We state our results in terms
of $\bm{\mathcal{L}}_1^s$. However, these results can be reformulated in terms
of the left and right eigenvectors of $\nLap_1$. In particular, if
$\nLap_1^s\bm u = \lambda \bm u$, then $\nLap_1\bm u_R = \lambda \bm u_R$ with
$\bm u_R = \bm D_2^{1/2}\bm{u}$, and similarly $\bm u_L^\top=\bm u^\top \bm
D_2^{-1/2}$ for the left eigenvectors of $\nLap_1$.
\begin{theorem}
    Consider the matrices 
    \begin{equation}
        \bm{G}_1 = \bm{D}_1^{-1/2}\bm{B}_1^{}\bm{D}_2^{}\bm{B}_1^\top\bm{D}_1^{-1/2} 
         \quad \text{and} \quad 
         \bm{G}_2 = \bm{D}_3^{1/2}\bm{B}_2^\top \bm{D}_2^{-1}\bm{B}_2^{}\bm{D}_3^{1/2}.
    \end{equation}
    Then the following statements hold:
    \begin{enumerate}
        \item Every eigenvector $\bm u$ of $\bm{G}_1$ with eigenvalue $\lambda$ has a corresponding eigenvector of $\nLap_1^{s}$ of the form $\bm v= \bm{D}_2^{1/2}\bm{B}_1^\top\bm{D}_1^{-1/2}\bm u$ with eigenvalue $\lambda$.
        \item Every eigenvector $\bm u$ of $\bm{G}_2$ with eigenvalue $\lambda$ has a corresponding eigenvector of $\nLap_1^{s}$ of the form $\bm v= \bm{D}_2^{-1/2}\bm{B}_2^{}\bm{D}_3^{1/2}\bm u$ with eigenvalue $\lambda$.
    \end{enumerate}
\end{theorem}
The matrix $\bm{G}_1\in \mathbb{R}^{n_0\times n_0}$ involves only lower adjacent couplings of the edges and 
has the form of a weighted graph Laplacian (recall that the the standard graph Laplacian is $\bm L_0 = \bm B_1 \bm B_1^T$).
Similarly, $\bm{G}_2 \in \mathbb{R}^{n_2\times n_2}$ is completely determined by weighted upper adjacent couplings of edges.
In both cases, the corresponding eigenvectors of $\nLap_1^s$ can be understood in terms of the Hodge decomposition above (\cref{eq:Hodge_weighted}).
In particular, the above result shows that eigenvectors of $\bm{G}_1$ or $\bm{G}_2$
with near-zero eigenvalues correspond to eigenvectors of $\nLap_1^s$ with
near-zero eigenvalues (i.e., nearly harmonic functions).

\begin{proof}
    Recall that the $\nLap_1^s$ is defined as:
    \begin{equation}
         \nLap_1^s= 
         \bm{D}_2^{1/2}\bm{B}_1^\top\bm{D}_1^{-1}\bm{B}_1^{} \bm{D}_2^{1/2} 
         + \bm{D}_2^{-1/2}\bm{B}_2^{}\bm{D}_3\bm{B}_2^\top \bm{D}_2^{-1/2},
    \end{equation}
    Let $\bm{G}_1\bm u= \lambda \bm u$ and $\bm v= \bm{D}_2^{1/2}\bm{B}_1^\top\bm{D}_1^{-1/2}\bm u$.
    Then
    \begin{align*}
        \nLap_1^s\bm v &= \bm{D}_2^{1/2}\bm{B}_1^\top\bm{D}_1^{-1}\bm{B}_1^{} \bm{D}_2^{1/2}\bm v = \bm{D}_2^{1/2}\bm{B}_1^\top\bm{D}_1^{-1}\bm{B}_1^{} \bm{D}_2^{}\bm{B}_1^\top\bm{D}_1^{-1/2}\bm u\\
                       &= \bm{D}_2^{1/2}\bm{B}_1^\top\bm{D}_1^{-1/2}\bm{G}_1\bm u = \lambda \bm{D}_2^{1/2}\bm{B}_1^\top\bm{D}_1^{-1/2}\bm u =\lambda \bm v,
    \end{align*}
    The proof for the second statement is analogous.
\end{proof}

\section{Constructing simplicial complexes and computation}
\label{sec:computation_and_constructions}
Before delving into applications, we first discuss computational aspects of our
diffusion framework.
The computational cost for higher-order interactions is larger 
compared to graph-based techniques, but often not prohibitively so.
Simpliclal complexes (SCs) built from data typically induce sparse (co-)boundary matrices.
Once the SC is constructed, the computations boil down to sparse matrix-vector
products, where the sparsity is linear in the number of elements of the
SC.
This holds when we move to even higher-order Hodge Laplacians, although the
size of the SC can grow.

\subsection{Constructing simplicial complexes from data}
We have thus far assumed that we are given an SC.
However, an SC $\SC$ is typically derived from data in some manner. 
There are several ways this is done in practice:
\begin{enumerate}
\item The original data is a collection of sets $\mathcal{S}$, and we induce $\SC$
  from elements of $\mathcal{S}$. For example, $\SC$ could be induced by all sets in $\mathcal{S}$
  with cardinality at most three. Alternatively, we could also
  include size-3 subsets of sets in $\mathcal{S}$ containing more than three elements. This
  practice has been used when, for example, the elements of $\mathcal{S}$ are sets of authors on
  scientific publications~\cite{Patania-2017-shape} or sets of tags annotating
  questions on Stack Overflow~\cite{Benson2018}.

\item The original data is a point cloud in a metric space, and we use geometric
  methods to construct, e.g., a Vietoris--Rips complex or \v{C}ech complex.
  This is standard practice in persistent homology~\cite{carlsson2009topology},
  where the data might be a time series of synaptic firings in a brain~\cite{Giusti-2016-neural}
  or a set of images~\cite{Lee-2003-images}.

\item The data is a graph, and $\SC$ is the clique complex, where
  2-simplices are the 3-cliques in the graph. The 0- and 1-simplices
  are given by the graph~\cite{Jiang2011,Lim-2015-hodge}.
\end{enumerate}
The first case is a ``top-down'' construction of the SC, while
the second case is a ``bottom-up'' approach. The third case is somewhere in
between---the graph structure imposes the $1$-skeleton, and the $1$-skeleton
contains all of the information for the SC.

The computational complexity of constructing the SC differs in each case. 
When the data is already a collection of sets, one might only need to process sets one-by-one. 
For the case where the data is a point cloud, finding fast algorithms for constructing SCs (or sequences of SCs) is an active area
of research~\cite{Chambers-2008-testing,Zomorodian-2010-fast,deSilva-2011-dualities,Chen-2011-twist,henselman2016matroid,Otter-2017-roadmap}.
Likewise, enumerating triangles in a graph to construct a clique complex is a well-studied
problem~\cite{Chiba-1985-subgraphs,Latapy-2008-triangles}. In the worst case,
the running time is $O(n_1^{3/2})$, where $n_1$ is the number of edges (1-simplices). 
However, practical algorithms are typically faster on real-world
graph data exhibiting common structural properties~\cite{Latapy-2008-triangles,Berry-2014-simple}.

\subsection{Solving $\mathcal{L}_1$ systems and the cost of matrix-vector multiplication}
Now suppose we have constructed an SC with a maximum simplex size
of three. Computationally, our applications will solve systems and compute
eigenvectors of matrices that involve the normalized Hodge Laplacian. In typical
data applications (including the ones we study later), the SC
induces a sparsely-representatable normalized Hodge Laplacian. Often,
approximate solutions are sufficient for data analysis, so iterative methods are
a natural choice in our computations. The driving factor in the running time of
the computation (ignoring issues of conditioning) is the cost of a matrix-vector
product of the normalized Hodge Laplacian.

Recall that the normalized Hodge 1-Laplacian is
\[
\bm{\mathcal{L}}_1 = \bm{D}_2\bm{B}_1^\top\bm{D}_1^{-1}\bm{B}_1^{} + \bm{B}_2^{}\bm{D}_3\bm{B}_2^\top \bm{D}_2^{-1},
\]
where the matrices $\bm{D}_i$ are simple diagonal matrices that can be directly computed from $\bm{B}_1$ and $\bm{B}_2$ with a single matrix vector product.
Suppose we have constructed $\bm{B}_1$ and $\bm{B}_2$.
Then the cost of applying the matrix $\bm{\mathcal{L}}_1$ to a vector involves a matrix vector product with the diagonal matrices $\bm{D}_i$ (a simple scaling), and the matrices
$\bm{B}_2^\top$, $\bm{B}_2$, $\bm{B}_1$, and $\bm{B}_1^\top$.
These incidence matrices have $O(n_1 + n_2)$, $O(n_1 + n_2)$, $O(n_0 + n_1)$, and $O(n_0 + n_1)$ nonzeros\footnote{We use $O(\cdot)$ notation here to emphasize that the number of nonzero entries scales linearly in the size of the simplices, even though the exact number of nonzeros could be given. For instance the matrix $\bm B_1$ has exactly $2n_1$ nonzero entries.}
, respectively, where $n_0$ is the number of 0-simplices (nodes),
$n_1$ is the number of 1-simplices (edges), and
$n_2$ is the number of 2-simplices (filled triangles).
Putting everything together, we can compute a matrix-vector product of the normalized Hodge Laplacian in 
time $O(n_0 + n_1 + n_2) = O(\lvert \SC \rvert)$, i.e., linear time in the size of the data. 
We note that there is also a growing literature on fast solvers for Laplacian systems~\cite{vishnoi2013lx} 
with some results for Hodge 1-Laplacians~\cite{cohen2014solving}.

The other major computational component for applications is the Hodge decomposition of edge flows, which we discuss next.

\subsection{Computation and interpretation of the Hodge Decomposition for edge flows}
\label{sec:hodge_background}

As discussed in \Cref{sec:hodge_background}, the Hodge decomposition is an
orthogonal decomposition of a vector space. We use the normalized decomposition
in \cref{eq:Hodge_weighted} to provide additional insights in our upcoming
applications. The normalized Hodge decomposition of a vector $\bm{c} \in
\mathbb{R}^{n_1}$ (an edge flow) is:
\begin{align}
  \bm{c} = \bm{g} \oplus \bm{r} \oplus \bm{h}, & \quad \text{ where }\;\; 
  \bm{g} = \bm{D}_2^{1/2}\bm{B}_1^\top\bm{p},\quad \bm{r} = \bm{D}_2^{-1/2}\bm{B}_2\bm{w},\quad 
  \bm{\mathcal{L}}_1^s\bm{h} = 0. \label{eqn:L1_decomp}
\end{align}

Since the decomposition is orthogonal, computing the decomposition boils down to
solving least squares problems:
\begin{align}\label{eqn:ls_hodge}
  \min_{\bm{p}} \| \bm{D}_2^{1/2}\bm{B}_1^\top\bm{p} - \bm{c} \|_2, \qquad
  \min_{\bm{w}} \| \bm{D}_2^{-1/2}\bm{B}_2\bm{w} - \bm{c} \|_2.
\end{align}
One must exercise a bit of caution here.
Although least squares problems are typically overdetermined, $\bm{D}_2^{1/2}\bm{B}_1^\top$ and $\bm{D}_2^{-1/2}\bm{B}_2$ are
rank-deficient exactly when $\text{ker} (\bm{\mathcal{L}}_1^s)$ is non-trivial (or, equivalently, 
$\text{ker} (\bm L_1)$ is non-trivial), i.e., when the SC has a non-trivial first cohomology group $\mathcal H_1$. 
For our purposes, we do not actually need to recover $\bm{p}$ or $\bm{w}$; we only need the residuals of the least squares problems. Let
$\bm{e}_p = \bm{D}_2^{1/2}\bm{B}_1^\top\bm{p}^* - \bm{c}$ and
$\bm{e}_w = \bm{D}_2^{-1/2}\bm{B}_2\bm{w}^* - \bm{c}$
be the residual error vectors for the least squares problems in
\cref{eqn:ls_hodge} (with minimizers $\bm{p}^*$ and $\bm{w}^*$). Then the Hodge
decomposition is given as follows:
\begin{align}
  \bm{g} = \bm{e}_p + \bm{c} = \bm{D}_2^{1/2}\bm{B}_1^\top\bm{p}^*, & &
  \bm{r} = \bm{e}_{w} + \bm{c} =\bm{D}_2^{-1/2}\bm{B}_2\bm{w}^*, & &
  \bm{h} = \bm{c} - \bm{g} - \bm{r}.
\end{align}

As discussed above, both $\bm{B}_1$ and $\bm{B}_2$ are sparse, and an
approximate solution is often satisfactory. Thus, appropriate numerical methods
for minimum-length linear least squares problems are iterative solvers such as
LSQR~\cite{Paige-1982-LSQR} and LSMR~\cite{Fong-2011-LSMR}, which produce
sequences of residual error vectors. The running time and computational
complexity of these algorithms is largely driven by the sparsity of the
matrices. If the SC is small enough, then one could first compute the
Moore--Penrose pseudoinverses and then compute the projections; in this case,
the computational complexity is dominated by the cost of computing the
pseudoinverse, which is $O(n_1n_0^2 + n_2n_1^2)$.

\begin{figure}[tb!]
 \centering
 \includegraphics[width=\columnwidth]{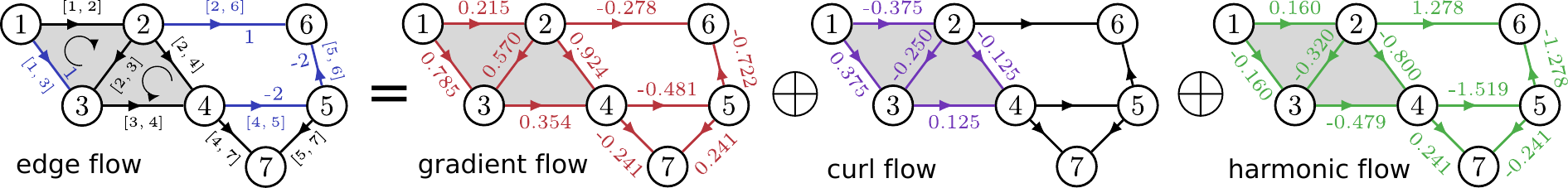}
 \caption{Standard (unnormalized/unweighted) Hodge decomposition
   (\cref{eq:Hodge}) of the edge flow in the example from
   \cref{fig:schematic}. The edge flow can be decomposed into orthogonal
   gradient, curl, and harmonic flows. Gradient flows around cycles sum to zero;
   curl flows are comprised of net flows around 2-simplices; and harmonic flows
   sum to zero around 2-simplices but are non-zero along longer
   cycles. Decomposing edge flows with the Hodge decomposition provides
   additional insights for data analysis applications. While the unnormalized decomposition
   is illustrative, in practice, we use the normalized Hodge decomposition (\cref{eq:Hodge_weighted}).}
 \label{fig:hodge_decomp}
\end{figure}

The components of the Hodge decomposition for edge flows are related to notions
from vector calculus~\cite{Lim-2015-hodge}.
The vector $\bm{g}$ is the projection of $\bm{c}$ into
$\text{im}(\bm{D}_2^{1/2}\bm{B}_1^\top)$, which is a weighted cut-space of the
edges~\cite{Schaub2014,Guattery1998,Godsil2013}, i.e., the linear combinations
of weighted edge vectors that disconnect the network. Equivalently, $\bm{g}$ is
a weighted gradient flow---the ``unweighted'' flow $\bm{D}_2^{-1/2} \bm{g}$ has
no cyclic component, meaning that the sum of its flow along any cyclic path in
the complex is zero, taking into account the orientations of the edges
(\cref{fig:hodge_decomp}, second from left).
The vector $\bm{r}$ is the projection of $\bm{c}$ into
$\text{im}(\bm{D}_2^{-1/2}\bm{B}_2)$, which consists of all weighted flows that
can be composed of local circulations along any 3-node simplex, i.e., weighted
circulations around filled triangles (the unweighted version is in
\cref{fig:hodge_decomp}, second from right). Indeed, the operator $\bm{B}_2$ is
a discrete analog of the familiar notion of a curl in vector
calculus~\cite{Lim-2015-hodge}. A high projection into the curl subspace thus
corresponds to a flow that is mostly composed of local circulations.
Finally, the harmonic component $\bm{h} \in \text{ker}(\nLap_1^s)$ corresponds
to a weighted version of a global circulation that does not sum to zero around
every cyclic path but is also inexpressible as a linear combination of curl
flows (the unweighted case is in \cref{fig:hodge_decomp}, far right). A flow
with a high-projection into the harmonic subspace is thus associated with global
cycles within the edge-space that can be directly related to the homology of the
SC.

\section{Application I: edge flow and trajectory embeddings}\label{sec:appplication_embedding}
The graph Laplacian with its connection to diffusion processes, harmonic analysis, and algebraic topology has been employed in many learning tasks, including manifold learning, dimensionality reduction, graph clustering, and graph signal processing~\cite{belkin2002laplacian,Coifman2006,Masuda2017,Ortega2018}.
Underpinning these methods is the spectral structure of the Laplacian.
Eigenvectors associated with $0$ eigenvalues are associated with the $0$th homology group of the graph, corresponding to connected components.
Eigenvectors with eigenvalues close to zero correspond to nearly disconnected components (clusters), as can be quantified by the celebrated Cheeger inequality.
By assessing the spectral properties of the graph Laplacian we can thus obtain an approximate notion of the topology, which embodies 
many spectral embedding and clustering techniques~\cite{belkin2002laplacian,Coifman2006}.
In the following, we translate these ideas to the context of the Hodge 1-Laplacian, by considering 
spectral embeddings of edges and flows.

\subsection{Synthetic data example}
We introduce our ideas with a synthetic example.
Consider a flow $\bm f$ defined on the edges of a SC as depicted in \cref{fig:flow_embedding_A}.
The underlying SC was constructed by (i) drawing $400$ random points in the unit square; (ii) generating a triangular lattice via Delauney triangulation; (iii) eliminating edges inside two predefined regions; and (iv) defining all triangles to be faces.
As depicted in \cref{fig:flow_embedding_A}, the SC $\SC$ has two ``holes.''
Accordingly, the (normalized) Hodge Laplacian has exactly two zero eigenvalues.
As depicted in \Cref{fig:flow_embedding_B,fig:flow_embedding_C}, these eigenvalues can be associated to two harmonic functions $\bm h_1$ and $\bm h_2$ that encircle the two holes in the SC.
To avoid differentiating between left and right eigenvectors, we use the symmetrized Laplacian $\bm{\mathcal{L}}_1^s$
here and in the remainder of this section, although all results can be translated to $\bm{\mathcal{L}}$ 
using the spectral relationships from \Cref{sec:norm_hodge_lap}.

\begin{figure*}[tb!]
    \centering
    \phantomsubfigure{fig:flow_embedding_A}
    \phantomsubfigure{fig:flow_embedding_B}
    \phantomsubfigure{fig:flow_embedding_C}
    \phantomsubfigure{fig:flow_embedding_D}
    \phantomsubfigure{fig:flow_embedding_E}
    \phantomsubfigure{fig:flow_embedding_F}
    \includegraphics[width=\textwidth]{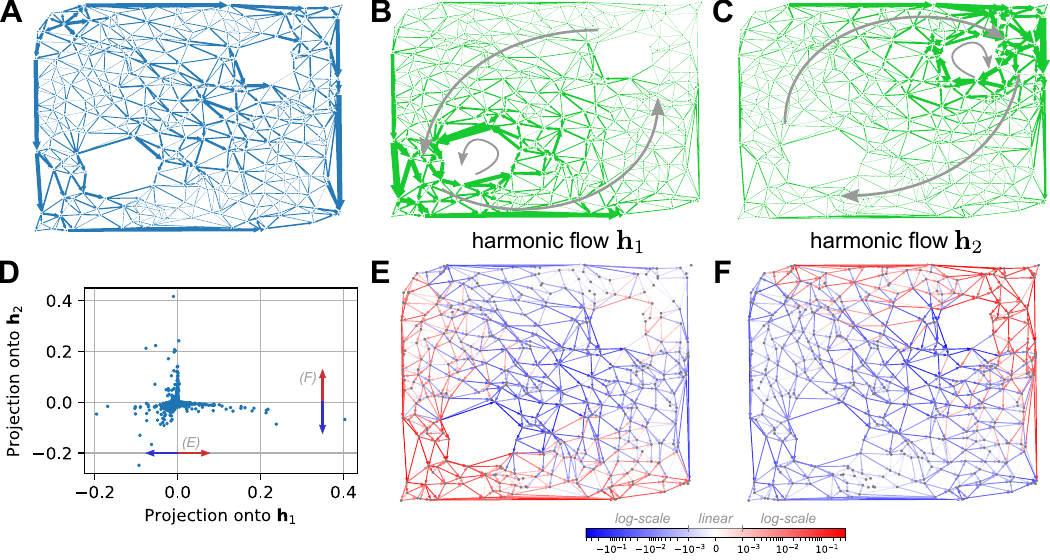}
    \caption{\textbf{Embedding of edge-flows.}
        \textbf{A} A flow on a simplicial complex with two ``holes,'' constructed as described in the text. Arrows indicate the direction of the flow on each edge, the magnitude is proportional to the width of the edge.
        \textbf{B--C} Harmonic functions $\bm{h}_1$ and $\bm{h}_2$ of the symmetric normalized Hodge 1-Laplacian $\nLap_1^s$ of the underlying SC. Edge directions correspond to the orientation induced by each harmonic function. 
        Gray arrows indicate how the harmonic flows encircle the two holes.
        \textbf{D} Projection of each edge flow $f_{[i,j]}$ (depicted in~\Cref{fig:flow_embedding_A}) onto the harmonic functions.
        \textbf{E--F} Projection of the edge flow onto the harmonic functions $\bm{h}_1$ (\Cref{fig:flow_embedding_E}) and $\bm{h}_2$ (\Cref{fig:flow_embedding_F}).
        Red indicates a positive projection, blue a negative projection. 
        The arrow direction is the same as in (A).
    }
    \label{fig:flow_embedding}
\end{figure*}

\paragraph{Edge flow embeddings}
Following ideas from Laplacian eigenmaps~\cite{belkin2002laplacian} and diffusion maps~\cite{Coifman2006} for the embedding of nodes of a graph, consider the spectral decomposition of the symmetric normalized 
Hodge 1-Laplacian $\nLap_1^s = \bm U \bm \Lambda \bm U^\top$.
Here, $\bm U = (\bm u_1, \ldots,\bm u_{n_1})$ is the matrix containing the eigenvectors of $\nLap_1^s$ and $\bm \Lambda = \text{diag}(\lambda_1,\ldots,\lambda_{n_1})$ is the diagonal matrix of eigenvalues, where we assume that the eigenvalues have been ordered in increasing magnitude $0 \le \lambda_1 \le \lambda_2 \le \ldots \le \lambda_{n_1}$.
For a Laplacian $\nLap_1^s$ with $k$ zero eigenvalues, we define $\bm H := (\bm u_1,\ldots,\bm u_k)$ to be the matrix collecting all the harmonic functions associated to $\nLap_1^s$.\footnote{Note that if $\lambda=0$ is a degenerate eigenvalue, the eigenvectors are only defined up to a unitary transformation and hence not unique; however, the subspace of harmonic functions is unique.}

Let us denote the indicator vector of a positive flow on the positively oriented
edge $e=[i,j]$ by $\bm e$, i.e., $\bm e_{[i,j]} = 1$ and $0$ otherwise.
We now define the \emph{harmonic embedding} of an edge $e$ via the mapping 
\begin{equation}
    e \mapsto \bm{l}_e= \bm H^\top \bm e \in \mathbb{R}^{k}.
\end{equation}
The embedding measures how a unit flow along the oriented edge $[i,j]$ projects into the harmonic subspace, i.e., how much it contributes to the global circular flows represented by the harmonic functions in terms of an inner product.
The reference orientation of the edge is important in that a positive projection coordinate indicates that the edge is aligned with the harmonic function; a negative coordinate signifies that the orientations are not aligned.

\Cref{fig:flow_embedding_A} shows a flow $\bm f$ on the SC.
For each edge $e = [i, j]$, we construct the weighted indicator vector $\bm f_{[i,j]} = f([i,j]) \bm e$ whose value is
the amount of flow on the edge relative to the chosen reference orientation.
We then compute the projection of this vector into the harmonic subspace $\bm{l}_f= \bm H^\top \bm f_{[i,j]}$.
\Cref{fig:flow_embedding_D} shows the embedding for each edge in the SC
with respect to the two harmonic flows of the SC (\Cref{fig:flow_embedding_B,fig:flow_embedding_C}).
Edges with a positive projection onto $\bm{h}_1 = \bm{u}_1$ are primarily aligned with flows that encircle the lower hole in the complex in the counterclockwise direction (red edges) and edges with a negative projection (blue edges) contribute to clockwise rotations (\Cref{fig:flow_embedding_E}).
An analogous argument holds, \emph{mutatis mutandis}, for the second coordinate corresponding to the projection onto $\bm{h}_2$ (\Cref{fig:flow_embedding_F}).
We emphasize that our explanation is geometric,
but the extracted features derive solely from the topological information encoded in the SC. 

We could use embedding coordinates other than the harmonic ones.
For instance, we may choose to project into the curl subspace or the gradient subspace, thereby revealing complementary information, such as how much the edge is aligned with the cut-space.
Alternatively, similar to Laplacian eigenmaps~\cite{belkin2002laplacian}, we may project onto the first $k'$ 
eigenvectors of $\nLap_1^s$, where $k'$ may be different from the size of the harmonic subspace.
Such a procedure would also account for contributions of an edge into parts of the gradient and the curl subspace, namely those associated with small eigenvalues (i.e., near-harmonic functions~\cite{Muhammad2006}).

\paragraph{Trajectory embeddings}
Inspired by Gosh et al.~\cite{ghosh2018topological}, we now consider
the embedding of ``trajectories'' of flow vectors $\bm f$ defined on contiguous edges
into the harmonic space.
\Cref{fig:trajectory_embedding_A} displays the same SC as above with a set of 9 trajectories.
Each of these trajectories is represented by a vector $\bm{f}$ with entries $f_{[i,j]} = 1$ if $[i,j]$ is part of the trajectory; $f_{[i,j]} = -1$ if $[j,i]$ is part of the trajectory; and $f_{[i, j]} = 0$ otherwise.
\Cref{fig:trajectory_embedding_B} shows the embedding of the complete flow vector along with its ``temporal evolution'' (dashed lines) in the embedding space, where we update the embedding one edge at a time, leading to a trajectory in embedding space.
The embedding differentiates the topological properties of the trajectories.
For example, the red, orange and green trajectory traverse the lower left obstacle aligned with $\bm{h}_1$.
Consequently, they have a similar embedding.
Similarly the brown, pink and violet trajectories are clustered together in the embedding space, as are the cyan, grey and olive green trajectory, reflecting their similarity in terms of their projection onto the harmonic subspace.

\begin{figure*}[tb!]
    \centering
    \phantomsubfigure{fig:trajectory_embedding_A}
    \phantomsubfigure{fig:trajectory_embedding_B}
    \includegraphics[width=0.9\textwidth]{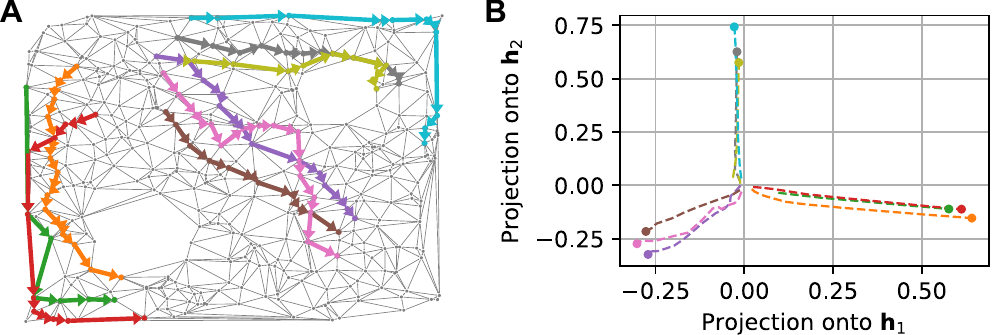}
    \caption{\textbf{Embedding of trajectories.}
        \textbf{A} A set of 9 trajectories, defined on the SC in \cref{fig:flow_embedding}. 
        Arrows indicate the direction of the flow on each edge.
        \textbf{B} Projection of each of the shown trajectories onto the harmonic eigenvectors of $\bm{\mathcal{L}}_1^s$.}    \label{fig:trajectory_embedding}
\end{figure*}

\subsection{Analysis of ocean drifter data}
The above idea of trajectory embeddings can be used for a number of tasks in data analysis, similar to node embeddings~\cite{belkin2002laplacian,Coifman2006}.
This includes clustering trajectories according to their relative position in the embedding space, definitions of similarity scores between trajectories (even if they have different lengths), and filtering noisy trajectories~\cite{schaub2018flow}.

The type of data for which these embeddings can be useful include any type of flow flows on some discrete (or discretized) domain.
For instance, this could be trajectories measured in two-dimensional physical space.
To construct a simplicial complex from such data we can, e.g., discretize such a flow on a hexagonal grid and choose each hexagonal cell to correspond to a node. An edge is then defined if the number of trajectories crossing from one cell to another exceeds a certain threshold; and a face is defined if sufficiently many trajectories pass through all three neighboring cells.
A missing face thus corresponds to an obstacle through which little flow passes.

This style of trajectory analysis was recently used to analyze mobility data with differential forms~\cite{ghosh2018topological}.
In contrast to this work, our formalism depends only on the construction of the SC.
Therefore, our methodology is not limited to planar SCs, even though we use planar examples for easy visualization.
Indeed, trajectory data collected in applications often has no explicit geometry, but may be understood as a sequence of nodes on an SC (or graph), e.g., moving from one Web page to another~\cite{Benson-2016-sequences}.
The construction of SCs from such data is a modeling question, 
where edges and faces can be chosen as a function of the observed flow pattern.
For example, we may create a face from a triangle of nodes if there is enough cyclic flow within this triangle,
or we may choose to assign weights to faces to regularize certain aspects of the data~\cite[Chapter 4]{grady2010discrete}.
The success of such methods will be context-dependent and contingent on whether the SC model provides a useful interpretation of the data.

\begin{figure*}[tb!]
    \centering
    \phantomsubfigure{fig:drifters_embedding_A}
    \phantomsubfigure{fig:drifters_embedding_B}
    \phantomsubfigure{fig:drifters_embedding_C}
    \includegraphics[width=\textwidth]{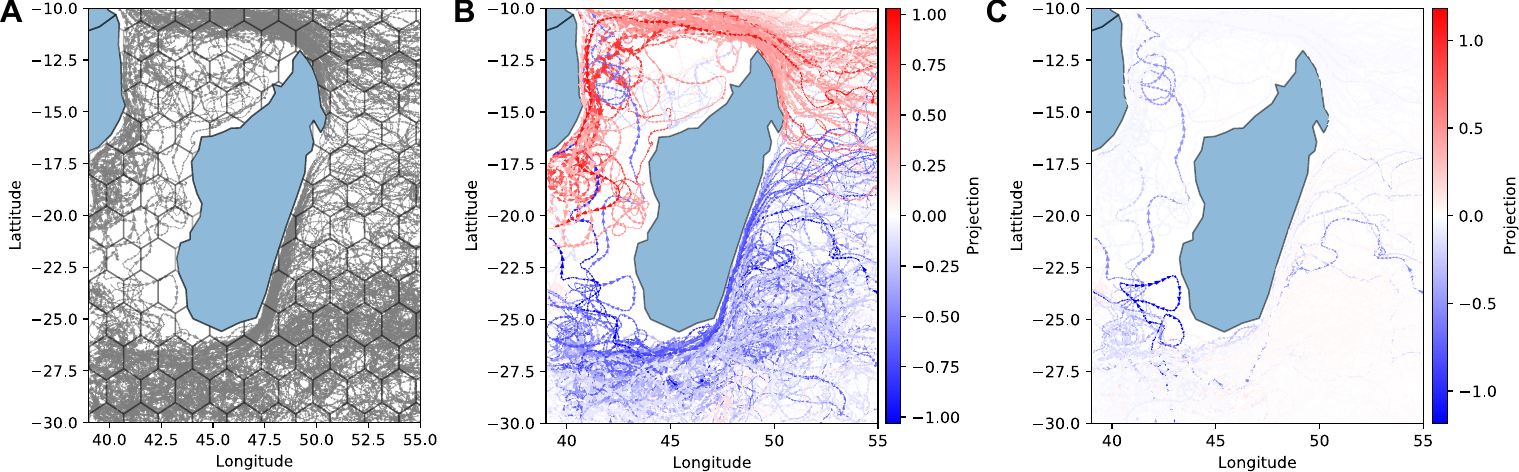}
    \caption{\textbf{Harmonic embeddings of ocean drifter trajectories.}
        \textbf{A} Visualization of buoy trajectory data around Madagascar. 
        Trajectories are discretized by a hexagonal grid shown in the background. 
        \textbf{B} Projection of the discretized trajectories onto the first harmonic flow. 
        Trajectories encircling Madagascar in a clockwise direction have a strongly negative projection (blue), and 
        trajectories encircling in a counter-clockwise direction a positive projection (red).
        \textbf{C} Projection of the individual trajectories onto the second harmonic flow,
        which corresponds to an almost local circulation near Southwest Madagascar.
        Only a small number of trajectories have a large projection onto this flow.
    }
    \label{fig:trajectory_drifters}
\end{figure*}

We now apply our harmonic embedding technique for analyzing data from the Global Ocean Drifter Program available at the AOML/NOAA Drifter Data Assembly Center.\footnote{\url{http://www.aoml.noaa.gov/envids/gld/}}
This data has been analyzed for detecting Lagrangian coherent structures~\cite{froyland2015rough}, where it was shown that certain flow structures (ocean current) stay coherent over time. 
While the entire dataset spans several decades of measurements, 
we focus on data from Jan 2011--June 2018 and limit ourselves to buoys 
that have been active for at least 3 months within that time period.
We construct trajectories by considering the location information of every buoy every 12 hours.
As buoys may fail to record a position, there are trajectories with missing data. 
In these cases, we split the trajectories into multiple contiguous trajectories.
For our analysis, we examine trajectories around Madagascar with a latitude 
$y_\text{lat} \in [-30,-10]$, and longitude $x_\text{long} \in [39, 55]$ (\cref{fig:trajectory_drifters}). 
This results in $400$ total trajectories.

To construct a SC, we first transform the data into Euclidean coordinates via an area-preserving (Lambert) projection.
We discretize Euclidean space using a hexagonal grid as shown in \cref{fig:drifters_embedding_A}, with the width of the hexagon equal to $1.66^\circ$ (latitude).
Each hexagon corresponds to a node, and we add an edge between two such nodes if there is a nonzero net flow from one hexagon to its adjacent neighbors.
We consider all triangles (3-cliques) in this graph to be faces of the SC.
The Laplacian $\nLap_1^s$ of the resulting SC has a two-dimensional harmonic space.
Each dimension of this space corresponds to an ``obstacle'' of the flow.

Finally, we discretize each trajectory by rounding its positional coordinates to
the nearest hexagon and consider the resulting sequence of edges that the
trajectory traverses in the SC.
\Cref{fig:drifters_embedding_B,fig:drifters_embedding_C} show the results of
projecting each trajectory into the harmonic subspace of $\nLap_1^s$, where we
color the original trajectories according to the projection score.
\Cref{fig:drifters_embedding_B} shows that the first harmonic function captures
the effect of Madagascar as an island. Specifically, the first harmonic
function separates the south equatorial current arriving at East of Madagascar
into the (i) current flowing towards north of Madagascar and (ii) the East
Madagascar current which flows southwards. The second harmonic function
corresponds to a more localized feature (\cref{fig:drifters_embedding_C}) caused
by a loopy current near Southwest Madagascar. Accordingly, most trajectories
have small projection onto this space.
 
\section{Application II: PageRank on simplicial complexes}\label{sec:applicationPR}
Centrality measures, initially conceptualized to quantify the social power of
individuals within social
networks~\cite{Katz1953,Bonacich1987,Freeman1977,Freeman1978,Boldi2014}, are an
important network analysis tool.
For instance, centrality measures have been used to identify pivotal elements in
infrastructure networks~\cite{Albert2000,Callaway2000} and to target
critical nodes in epidemic spreading processes on
networks~\cite{Kitsak2010,Pastor-Satorras2015}.
One of the most widely adopted centrality measures is PageRank, which can be
interpreted in terms of a random walk on a graph~\cite{Brin2012}.
Initially introduced as a ranking mechanism for hyperlinked webpages, PageRank
has been extensively used and studied in other
contexts~\cite{Langville2004,Bianchini2005,Gleich2015}.
For example, (personalized) PageRank has been used for graph clustering,
community detection, and semi-supervised learning~\cite{Andersen2006,Gleich-2015-local,Gleich2015,Kloumann2017,Schaub2017}.

Analogous to the situation of graphs, we would like to assess the importance of certain
simplices within an SC, i.e., extend the theory of centrality measures for graphs.
One could use simple notions such as the degree of a simplex, 
but extending other notions of centrality to SCs is non-trivial~\cite{Estrada2018}.
Here we leverage the connection of PageRank to random walks to derive a PageRank
measure from the $\bm{\mathcal{L}_1}$ Laplacian that extracts the topological
importance of edges in a SC.

\subsection{Background: (personalized) PageRank on graphs}
Following Gleich~\cite{Gleich2015}, we adopt the following definition of PageRank.
\begin{definition}[PageRank on graphs~\cite{Gleich2015}]\label{def:PageRankGraph}
    Let $\bm{P}$ be a column-stochastic matrix, $\bm{\mu}$ a stochastic column
    vector with $\1^\top \bm{\mu} = 1$, and $\alpha \in (0,1)$
    a teleportation parameter. The PageRank vector $\bm\pi$ is the solution to
    the following linear system:
    \begin{equation}
        (\bm{I} - \alpha \bm{P})\bm\pi = (1-\alpha)\bm{\mu}.
    \end{equation}
\end{definition}
The PageRank vector $\bm\pi$ is the stationary distribution of a random walker
on a graph, who at each step makes transitions according to $\bm{P}$ with
probability $\alpha$, and with probability $1-\alpha$ teleports to a random node
according to the probability distribution $\bm{\mu}$. The factor $1-\alpha$
facilitates the random walk interpretation but is often omitted as it is simply
multiplicative scaling~\cite{Gleich2015}.

There are two common types of PageRank~\cite{Gleich2015}. In standard
PageRank~\cite{Brin2012}, the teleportation distribution is uniform ($\bm{\mu} = \1/n_0$) and
the PageRank vector $\bm\pi$ is used to rank nodes. In
``personalized'' PageRank, $\bm \mu$ is an indicator vector on a node $i$, so
with probability $1-\alpha$ we restart our random walk process at $i$. The
resulting PageRank vector $\bm\pi$ can be interpreted as the influence node $i$
exerts on others: the $j$th entry of $\bm\pi$ is large if $i$ is well connected
to $j$. One can thus find nodes tightly coupled to $i$, a feature that can be
employed for local community detection~\cite{Andersen2006,Kloumann2017} and
graph-based semi-supervised learning~\cite{Gleich-2015-local}.

\subsection{PageRank vectors on simplicial complexes}
To generalize PageRank to SCs, we consider the problem on the lifted random walk
matrix $\bm{\widehat{P}}$ from \cref{thm:stoch_lifting}:
\begin{equation}
    (\bm{I} - \alpha \bm{\widehat{P}}) \bm{\widehat\pi} = (1-\alpha)\bm{\mu}.
\end{equation}
As we are interested in the subspace of alternating functions within the lifted
space, we project the resulting PageRank vector back into
$\mathcal{C}_1$. Specifically, the vector $\bm{V}^\top \bm{\widehat\pi}$ gives
the simplicial PageRank value on oriented edges. We solve for this vector:
\begin{equation}
    (1-\alpha) \bm{V}^\top \bm{\mu} = \bm{V}^\top (\bm{I} - \alpha \bm{\widehat{P}}) \bm{\widehat\pi}
    = \left(\bm{V}^\top + \frac{\alpha}{2} \bm{\mathcal{L}}_1 \bm{V}^\top \right) \bm{\widehat\pi}
    = \left(\bm{I} + \frac{\alpha}{2} \bm{\mathcal{L}}_1\right) \bm{V}^\top \bm{\widehat\pi},
\end{equation}
where we used part (1) of \cref{thm:relations} in the second equality. Hence,
we can compute the projected PageRank vector using $\nLap_1$ and never have to
construct $\widehat{\bm P}$.

Equivalently, for $\beta = 2/\alpha$, we can write the system for $\bm{V}^\top \bm{\widehat\pi}$ as
\begin{equation}\label{eq:proj_SPR}
\bm{V}^\top \bm{\widehat\pi} =  (\beta - 2) (\beta \bm{I} + \bm{\mathcal{L}}_1)^{-1} \bm{V}^\top \bm{\mu}.
\end{equation}
It is insightful to compare the projected PageRank vector in \cref{eq:proj_SPR}
with the graph-based PageRank once more. Note that we can rewrite
\cref{def:PageRankGraph} as
\begin{flalign}\label{eq:PR_inv}
    \bm\pi &= (1-\alpha) (\bm{I} - \alpha \bm{P})^{-1}\bm{\mu}
    = \dfrac{(1-\alpha)}{\alpha}  \left(\dfrac{1}{\alpha} \bm{I} - \bm{P}\right)^{-1}\bm{\mu}
    = \beta_0 (\beta_0 \bm{I} + \nLap_0)^{-1}\bm{\mu},
\end{flalign}
where $\beta_0 = 1/\alpha - 1$. There is a striking similarity between
graph-based PageRank and the projected simplicial PageRank introduced here.
Indeed, \cref{eq:PR_inv} suggests a definition of simplicial PageRank similar to
\cref{eq:proj_SPR} on purely notational grounds. However, based on
\Cref{thm:stoch_lifting}, we know that there is a relationship to a random walk,
albeit in a lifted state space. While \cref{eq:proj_SPR} is interpretable in
terms of a random walk only for $\beta\in (2,\infty)$, the inverse
$(\beta\bm{I} +\nLap_1)^{-1}$
remains well-defined for smaller positive values of $\beta$, as
$(\beta\bm{I} +\nLap_1)$ is positive definite ($\nLap_1$ is similar
to the positive semi-definite matrix $\nLap_1^s$, so $\nLap_1$ is positive semi-definite).
We may thus choose to ignore the multiplicative
scaling $\beta-2$, leading to a generalized form of a PageRank vector. These
two variants are summarized in the following definition.
\begin{definition}[PageRank and generalized PageRank vectors for edges in SCs]
    Let $\SC$ be a simplicial complex with normalized Hodge 1-Laplacian
    $\nLap_1$, $\bm{x}$ be a vector of the form $\bm{x} = \bm{V}^\top\bm{\mu}$
    where $\bm{\mu} \in \mathbb{R}^{2n_1}$ is a probability vector, and $\beta\in(2,\infty)$. The
    PageRank vector $\bm{\pi}_1$ of the edges is then defined as the solution to
    the linear system
\begin{equation}
    \left(\beta \bm{I} + \nLap_1\right)\bm{\pi}_1 = (\beta -2)\bm{x}.
\end{equation}
The generalized PageRank vector (for any $\kappa \in \mathbb{R}^+$) is the solution of the linear system
\begin{equation}
    \left(\kappa \bm{I} + \nLap_1\right)\bm{\pi}_1^g = \bm{x}.
\end{equation}
\end{definition}

Since $\bm\pi_1$ corresponds to a projection of a diffusion in a lifted space,
the PageRank is effectively a smoothed out version of a distribution $\bm \mu$
in the lifted space $\mathcal D^1$ and thus measures how a starting distribution
$\bm \mu$ will be shaped by the structure of the SC. Similar to the graph case,
certain oriented edges will attract more probability in this process and will in
this sense be deemed more important. A key difference is the projection step,
which again emphasizes the importance of the orientation. The absolute value of
entry $[i,j]$ in $\bm \pi_1$ will be high if there is large difference in the
probability of being at edge $[i,j]$ in the lifted space as compared to $[j,i]$.

\paragraph{Simplicial PageRank interpretation as a filter for edge-space signals}
To provide further intuition for the above measures, we interpret $\bm{x}$ as a
signal defined in the edge-space, similar to our discussion on edge and
trajectory embeddings in \Cref{sec:appplication_embedding}. We discuss this
issue in terms of the generalized PageRank vector $\bm \pi_1^g$. For any such
signal $\bm x$, the PageRank vector corresponds to a transformation of this
signal $\bm x$ according to the \emph{generalized PageRank operator}
$\left(\kappa \bm{I} + \nLap_1\right)^{-1}$, for some $\kappa > 0$. To
understand how this multiplication acts on $\bm x$, write the spectral
decomposition of the Hodge Laplacian by
$\nLap_1 = \bm U_R \bm \Lambda \bm U_L^\top$,
where $\bm U_R$ and $\bm U_L$ are the matrices containing the right and left eigenvectors of $\nLap_1$, respectively.
From our discussion at the end of \Cref{sec:norm_hodge_lap}, we further know
that $\bm U_R = \bm D_2 \bm U_L$. The generalized PageRank operator acts as a
filter on the signal $\bm x$ by projecting $\bm x$ onto the spectral coordinates
of the Laplacian, scaling according to a shifted version of the spectrum (the eigenvalues of $\nLap_1$ are shifted by $\kappa$), and
projecting back on the SC:
\begin{equation}
    \bm{\pi}_1 
    = \left(\kappa \bm{I} + \nLap_1\right)^{-1}\bm{x} 
    = \bm U_R \left(\kappa \bm{I} + \bm \Lambda \right)^{-1} \bm U_L^\top \bm{x} 
    = \bm U_R \left(\kappa \bm{I} + \bm \Lambda \right)^{-1} \bm U_R^\top \bm D_2^{-1} \bm{x}.
\end{equation}
Thus, there is a close relationship to the embeddings discussed above. However,
in contrast to the harmonic embedding, all eigenvectors (modulated by their
eigenvalue) are taken into account in the PageRank filtering operation. This
type of filtering is analogous to the filtering of signals defined on the nodes
of a graph as considered in graph signal processing~\cite{Ortega2018}. The
PageRank operator may thus be understood as a filter for \emph{edge signals}.
Recent research provides additional filtering
perspective~\cite{schaub2018flow,barbarossa2016introduction} with connections to
discrete exterior calculus~\cite{grady2010discrete,jia2019graph}.

\paragraph{Edge orientations and personalized simplicial PageRank}
As the projection leading to simplicial PageRank corresponds to a difference of
two probabilities, there is no guarantee that a PageRank vector $\bm\pi_1$ has
only positive entries. In light of our previous discussions on edge
orientation, this again relates to the issue of defining reference orientation
for edges.

\begin{figure*}[tb!]
  \phantomsubfigure{fig:example_analysis_A}
  \phantomsubfigure{fig:example_analysis_B}
  \phantomsubfigure{fig:example_analysis_C}
  \phantomsubfigure{fig:example_analysis_D}
    \centering
    \includegraphics[width=\textwidth]{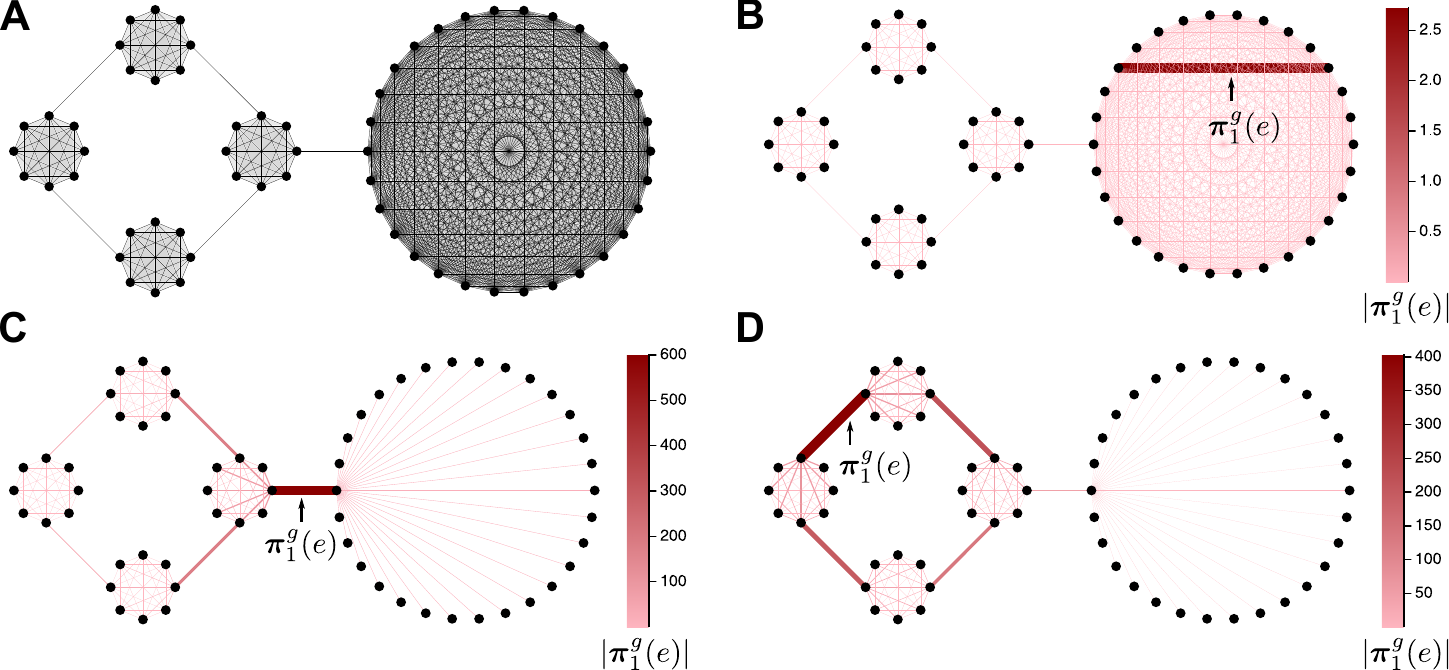}
    \caption{\textbf{Example analysis of a network using generalized PageRank vectors.} 
    \textbf{A} Simplicial complex consisting of 5 cliques. All triangular structures in the drawn graph skeleton correspond to 2-simplices, as indicated by the gray shading.
        \textbf{B--D} Generalized personalized PageRank vectors of three edges within different locations of the complex ($\kappa=0.001$).
     The color and the edge-width (the thicker the edge the larger the PageRank) indicate the magnitude of the component of the PageRank vector on the respective edge.
    The PageRank vector in (B) is localized and has small magnitude.
    The vector in (C) is almost localized but has large magnitude.
    Finally, the vector in (D) is not localized and has strong components around the cyclic structure.}
    \label{fig:example_analysis}
\end{figure*}

We can flip the reference orientation of an edge with a negative entry to obtain
a positive PageRank vector entry. The operation of flipping the orientation of an
edge can be cast as a gauge transformation or signature-similarity
transformation~\cite{Altafini2013} of the normalized Hodge 1-Laplacian. Let
$\bm\Theta_{n_1 \times n_1} = \text{diag}(\theta_1,\ldots,\theta_{n_1})$
be a diagonal matrix with $\theta_i= 1$ for all edges whose orientation is to
remain fixed, and $\theta_i=-1$ for those edges whose orientation we would like
to reverse. Then $\nLap_1' = \bm\Theta\nLap_1\bm\Theta$ describes the normalized
Hodge 1-Laplacian corresponding to the SC with the new edge orientations ($\bm \Theta$
defines a similarity transformation, so the spectral properties of
$\nLap_1$ are unaffected; we have merely changed the coordinate system in which
flows are measured). Using this transformation, we obtain the following
generalized PageRank vector:
\begin{equation}
    \bm\pi_1^g = \bm\Theta\left(\kappa \bm{I} + \nLap_1\right)^{-1}\bm\Theta \bm{x}.
\end{equation}
We claim that if $\bm{x}$ is an indicator vector, then $\bm\Theta$ can be chosen to make
$\bm\pi_1^g$ nonnegative. To see this, if $\bm{x}$ is an
indicator vector $\bm{e}_{[i,j]}$, it will pick out one column of
$\left(\kappa \bm{I} + \nLap_1\right)^{-1}$,
which through the action of $\bm\Theta$ can immediately be made non-negative on all entries
except at index $[i, j]$.
And since $\left(\kappa \bm{I} + \nLap_1\right)^{-1}$ is positive
definite, the diagonal entry at index $[i, j]$ will be positive.
Hence, just as with personalized PageRank on graphs, if $\bm{x}$ is an indicator vector
on edge $[i,j]$, we may interpret the absolute values of the entries
of $\bm\pi_1^g$ in terms of the influence edge $[i,j]$ exerts on the edges in
the simplicial complex. For data that does not induce a natural orientation of
the edges (in contrast to our previous examples) we may thus simply use the
element-wise absolute value of this ``personalized'' PageRank vector as an influence measure
between the edges.

While this trick of redefining the orientations of the edges also applies if
$\bm x$ is not an indicator vector, the situation is a bit more complicated.
Since the PageRank operator acts linearly on $\bm x$, we can decompose any
vector $\bm x$ into a weighted sum of unit vectors. For each of these unit
vectors we can assess the induced PageRank vector in terms of its absolute
value. However, due to differences in the sign patterns of the columns of
$\left(\kappa \bm{I} + \nLap_1\right)^{-1}$ the sum of these induced
``absolute'' (personalized) PageRank vectors is not the same as the absolute value of the
PageRank vector associated to $\bm x$, rendering the above interpretation in
terms of the influence of the individual edges difficult. In the following, we
will thus concentrate on personalized PageRank vectors with a teleportation
vector $\bm{x}$ localized on a particular edge $[i,j]$. We denote such a
personalized PageRank vector by $\bm\pi_1([i,j])$, or $\bm\pi_1^g([i,j])$ for
generalized PageRank.

\paragraph{Synthetic data example}
To illustrate our ideas, we again start with a synthetic example
(\cref{fig:example_analysis_A}). We construct a SC which consists of 4 groups
of 8 nodes (cliques) in a ring configuration, which is connected to a larger
clique of 30 nodes. All triangles (3-cliques) in the graph are faces in the SC.
\Cref{fig:example_analysis} visualizes three generalized personalized PageRank
vectors for $\kappa=0.001$. For the edge in \cref{fig:example_analysis_B}
located in the large clique, the generalized PageRank vector is effectively
localized on the edge itself. Moreover, the magnitudes of the entries of the
vector are small, indicating that its influence within the space of edges is
small. The edge in \cref{fig:example_analysis_C} corresponds to the ``bridge''
between the large clique and the ring of small cliques on the right hand side.
While the generalized PageRank vector is also concentrated on a few edges, the
magnitudes of its components are substantially larger (since we are using the
generalized PageRank, the entries can be substantially larger than 1). Finally,
for one of the cycle edges (\cref{fig:example_analysis_D}), we see a different
picture. In this case, the generalized PageRank vector is also of a large
magnitude, but most of the influence of this edge is concentrated on edges
around the cycle.

\subsection{Decomposing and aggregating simplicial PageRank vectors via the normalized Hodge decomposition}
\begin{table}[tb!]
    \centering
    \caption{Norms of components given by the normalized Hodge decomposition (\cref{eqn:L1_decomp}) for the generalized PageRank vectors in \cref{fig:example_analysis}.}
    \label{tab:norms}
    \begin{tabular}{lcccc}
        \toprule
        & & $\|\bm\pi_1^g\|_\text{grad}$ & $\|\bm\pi_1^g\|_\text{curl}$ & $\|\bm\pi_1^g\|_\text{harm}$\\
        \midrule
        ``Bulk'' edge   & \cref{fig:example_analysis_B} & 0.50 & 2.70 & 0 \\
        ``Bridge'' edge & \cref{fig:example_analysis_C} & 707.85 & 0 & 0 \\
        ``Cycle'' edge  & \cref{fig:example_analysis_D} & 265.14 & 0 & 518.15 \\
        \bottomrule
    \end{tabular}
\end{table}

By considering again the spectral properties of the normalized Hodge 1-Laplacian
$\nLap_1$, we see that this behavior is a consequence of the topological setup
of the SC. As discussed above, the eigenvectors associated to the zero
eigenvalues of $\nLap_1$ correspond to harmonic functions on the complex, which
are associated with the cycles in the graph not induced by $2$-simplices. As
the PageRank vector is computed via a shifted inverse of $\nLap_1$, edges that
have a significant projection into the null space of $\nLap_1$ will result in a
PageRank vector with a strong harmonic (cyclic) component. To better understand
the importance of the edges, it is insightful to consider the decomposition of
the PageRank vectors in terms of the normalized Hodge decomposition.

We can use the normalized Hodge decomposition from \cref{eqn:L1_decomp} to
compute the projections of the above computed PageRank vectors onto the
gradient, curl and harmonic subspaces. This gives a more nuanced picture of the
contributions of PageRank vectors (\cref{tab:norms}). The edge in
\cref{fig:example_analysis_B} actually has no harmonic part, and the curl and
gradient components largely cancel each other out, apart from the flow on the
edge itself. The edge in \cref{fig:example_analysis_C} lies in the
weighted cut-space. The norm of its corresponding gradient projection is large,
and its harmonic and curl projections are zero. Finally, the edge in
\cref{fig:example_analysis_D} is part of a harmonic cycle in the SC;
consequently, the norm of the PageRank vector projected into the harmonic
subspace is high.

\begin{figure}[tb!]
  \phantomsubfigure{fig:example_analysis2_A}
  \phantomsubfigure{fig:example_analysis2_B}
    \centering
    \includegraphics[]{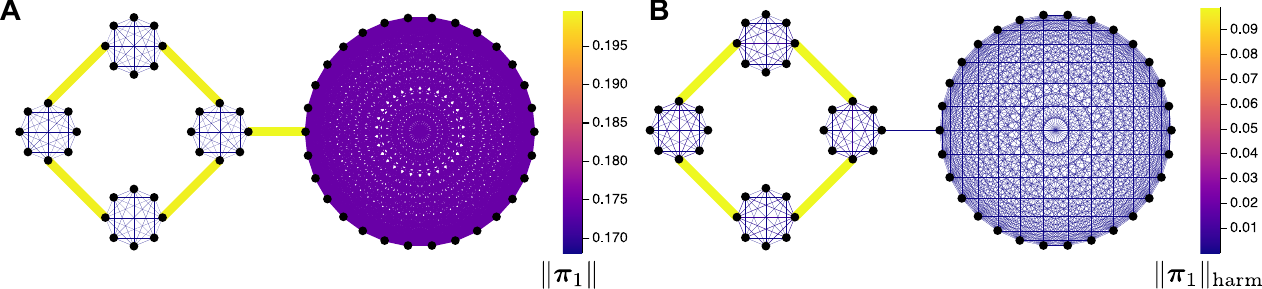}
    \caption{\textbf{Analyzing edges via norms of personalized PageRank vectors}. 
    \textbf{A} The 2-norm of the personalized PageRank vector associated to each edge. 
    \textbf{B} We can alternatively highlight the importance of an edge with respect to the homology 
    of the simplicial complex by considering the 2-norm of the projection of each PageRank vector
    into the harmonic space.}
    \label{fig:example_analysis2}
\end{figure}

Instead of decomposing the personalized PageRank vectors associated with each
edge, we can assess the importance of an edge within an SC in an aggregated
fashion with the 2-norm of the vector, or the 2-norm of any of the projections
given by the normalized Hodge decomposition. \Cref{fig:example_analysis2}
illustrates this procedure ($\beta=2.5$). In these diagrams, the value along an edge corresponds to the
2-norm of the personalized PageRank vector corresponding to the edge (\cref{fig:example_analysis2_A})
or to the 2-norm of the harmonic component of the personalized PageRank vector (\cref{fig:example_analysis2_B}).

Let us remark here that these are clearly not the only ways to extract a ``relevance score'' of the edges from the computed PageRank vectors and investigating other functions will be an interesting objective for future work.

\subsection{Analysis of political book co-purchasing data}
We now analyze a dataset of political book co-recommendations. The dataset
records the co-purchasing of 105 political books---as indicated by the
``customers who bought this book also bought these other books'' feature on
Amazon---around the time of the 2004 presidential election in the USA. The data
was collected by
Krebs~\cite{Krebs}\footnote{\url{http://vlado.fmf.uni-lj.si/pub/networks/data/cite/polBooks.paj}}
and subsequently analyzed by Newman~\cite{Newman2006}. Newman categorized the
books by hand according to their political alignment into 3 groups: ``liberal''
(43 books),``conservative'' (49 books), and books with bipartisan and centrist
views or no clear alignment (13 books), which we refer to as ``neutral.''
Although the data has been analyzed as a network, the edges represent frequent
co-purchasing of books by the same buyers, and thus the data has an implicit
simplicial structure (mulitple books are bought together, forming a simplex).
To construct an SC, we filled each triangle (3-clique) in the data to form a
2-simplex. \Cref{fig:pol_books_A} visualizes this SC, where gray shading
indicates the simplices. As already observed by Newman~\cite{Newman2006}, there
is a marked community structure, which is commensurate with the political
alignment of the books. There are two main clusters corresponding to liberal
and conservative books, along with a smaller group of books which act as bridges
between these two clusters.

\begin{figure*}[tb!]
  \phantomsubfigure{fig:pol_books_A}
  \phantomsubfigure{fig:pol_books_B}
  \phantomsubfigure{fig:pol_books_C}
    \centering
    \includegraphics[width=\textwidth]{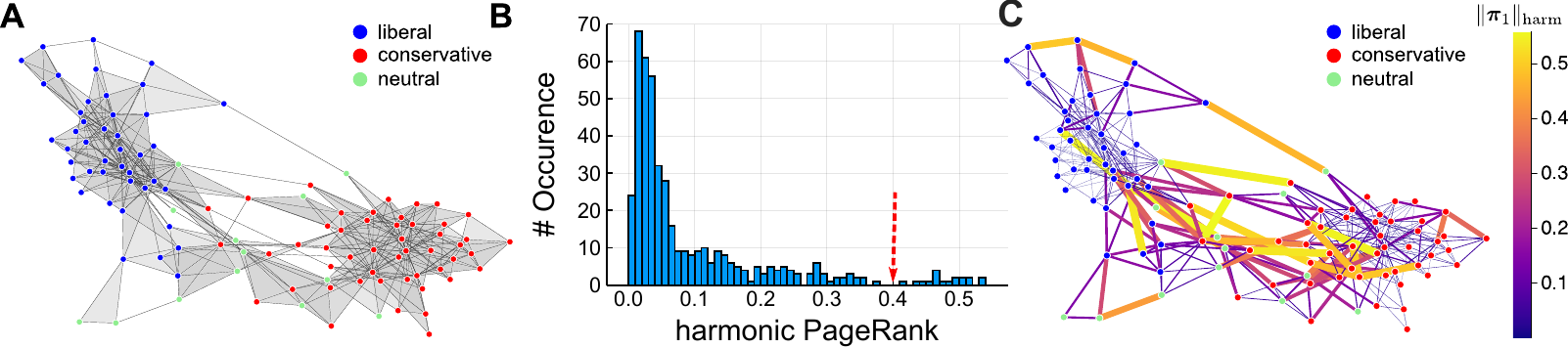}
    \caption{\textbf{Book co-purchasing analysis with Simplicial PageRank.} 
    \textbf{A} The co-purchasing of 105 political books---categorized
      as ``liberal,'' ``conservative,'' or ``neutral''---can be abstracted as a simplicial complex (gray areas are co-purchasing simplices). 
      \textbf{B} Harmonic PageRank values of edges. 
      We identify a set of 18 edges with a high harmonic PageRank value (threshold indicated by dashed red arrow). 
      \textbf{C} Visualization of all edges with their respective harmonic PageRank values.}
    \label{fig:pol_books}
\end{figure*}

To further analyze the importance of the co-purchasing with respect to the homology of the complex, 
we compute the ``harmonic PageRank values'' of the edges, which correspond to 
the 2-norm of the harmonic projection of the personalized PageRank score of each edge ($\beta = 2.5$).
Consistent with its importance for the homology, these values highlight edges that act as connectors around ``holes,'' which
indicate books that are never bought together as a set (\cref{fig:pol_books}). 
The ranking on edges induced by these harmonic PageRank values is robust with respect to the parameter $\beta$:
the Spearman rank correlation coefficient $\rho$ between the obtained PageRank
vectors for $\beta \in (2.05,2.67)$ (corresponding to a teleportation parameter
$\alpha \in (0.75,0.975))$ had a mean of $0.78$, providing evidence for the
consistency of the ranking.

\Cref{fig:pol_books_C} shows that even with the separation into
political clusters, there are still several edges with high harmonic PageRank
within each cluster.
Indeed, the edge with the largest aggregated harmonic PageRank corresponds to a
connection between two conservative books. To investigate this aspect further,
we plotted the histogram of the aggregated harmonic PageRank and identified a
tail of $18$ co-purchases with a harmonic PageRank greater than $0.4$
(\cref{fig:pol_books_B}). Out of the $18$ edges with the highest harmonic
score, $15$ are between books of the same category. Edges in the conservative
cluster with a high harmonic PageRank are more prevalent: $9$ of the $18$
highest aggregated harmonic centrality edges connect conservative books.
Thus, there appear to be ``gaps'' in the space of political opinions even within political clusters, indicating a fragmentation of
opinions.

\begin{figure}[tb!]
    \centering
    \includegraphics[]{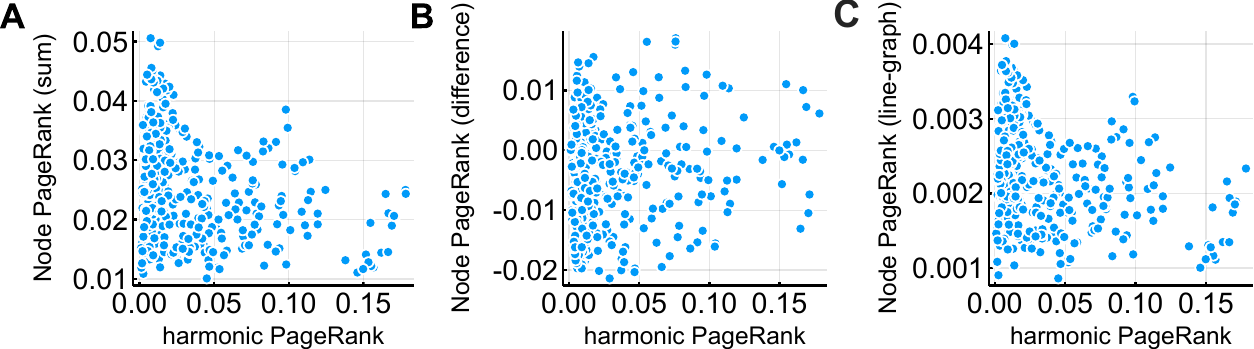}
    \caption{\textbf{Comparison of simplicial PageRank and graph-based PageRank.} The scatter plots compare
     the harmonic PageRank (see text) with various edge scores constructed from standard graph-based PageRank.
     \textbf{A} Edge scores are the sums of the graph-based PageRank scores of the incident nodes;
     the Spearman rank correlation is $\rho=0.04$.
     \textbf{B} Edge scores are the differences of the graph-based PageRank of the incident nodes;
     the Spearman rank correlation is $\rho=-0.08$.
    \textbf{C} Edge scores come from graph-based PageRank on the line graph;
    the Spearman rank correlation is $\rho = 0.08$.}    
    \label{fig:PR_correlation}
\end{figure}

Importantly, this information revealed by simplicial PageRank is complementary
to standard node-space PageRank (\cref{fig:PR_correlation}). We computed the
PageRank $\bm\pi$ of the nodes in the graph and assigned each edge (i) the sum or
(ii) the difference of the PageRank of its incident nodes. We also compute the
PageRank $\bm\pi$ from the line graph corresponding to the co-purchasing
network. In all cases, we use the standard teleportation parameter $\alpha =
0.85$. As can be seen from \cref{fig:PR_correlation}, all of these PageRank
scores on edges are essentially uncorrelated to the simplicial (harmonic)
PageRank, highlighting that we extract a different kind of topological
information from the data that what one could get with existing methodology.

\section{Discussion}\label{sec:discussion}
The connection between Markov chains, diffusions, and random walks on graphs
has led to successful analysis of complex systems and networks within applied mathematics.
However, as we are faced with increasingly complex and diverse datasets, 
some limitations of traditional graph-based models have come to the fore.
Researchers have thus employed richer modeling 
frameworks---such as multiplex networks, graphons, and simplicial complexes (SCs)---and 
investigated extensions of graph-based methods such as extensions of centrality 
measures~\cite{Sole-Ribalta2014,Avella-Medina2017,Estrada2018}.

SCs and other tools from algebraic and computational topology are promising in this pursuit of extending dyadic network models to account for polyadic interactions between groups of nodes, seeing success in a number of applications~\cite{Nanda2014,Chan2013,Petri2014,Giusti2015,Ghrist2005,DeSilva2007,Tahbaz-Salehi2010}.
However, topological tools have been mostly employed in static contexts, though some works have explored links to dynamical processes on network~\cite{Taylor2015}.
Here we introduced a normalized Hodge Laplacian operator that enables us to define a
diffusion processes on SCs in a principled manner, respecting the topological properties of the complex.
In particular, we have focused on diffusion processes in the edge-space with
applications in embedding edge trajectory data and simplicial extensions of PageRank for edge importance.

A number of avenues present themselves for future research.
There are many variants of random walks we may want to explore and compare to their graph-based counterparts~\cite{Masuda2017}.
In particular, a better understanding of how higher-order topological features impact on convergence, mixing, and other random walk properties is of interest in this context.
Another direction is to explore the translation of further random walk tools for SCs.
We focused on diffusion-based embeddings and centrality measures.
However, random walks have also been employed in many other learning tasks~\cite{Masuda2017}.
We expect that enriching such tools through the lens of SCs will be fruitful.

Finally, there is a rich history connecting random walks on graphs to connectivity and near $0$th-order homology, often formalized through ideas in spectral graph theory such as the discrete Cheeger inequality and isoperimetric inequality~\cite{Alon-1985-isoperimetric,Mihail-1989-conductance,Sinclair-1989-approximate,Lawler-1988-bounds,Chung-2007-heat,Kloster-2014-heat}.
For example, personalized PageRank---when viewed as an algorithm for local clustering for graphs---finds low conductance sets, formalized via a local Cheeger's inequality~\cite{Andersen2006}, which provides credence for our personalized polyadic walks.
A major open research direction is a clean generalization of these concepts to SCs.
There are a number of connections between the spectra of matrices (such as the Hodge Laplacian) associated 
with SCs and notions of expansion~\cite{Gundert-2012-laplacians,Doterrer-2012-coboundary,Steenbergen-2014-cheeger,Parzanchevski-2015-isoperimetric,Osting2017}.
However, some types of higher-order Cheeger-like inequalities are impossible due to torsion that may be present in the homology of SCs that is not present in graphs~\cite{Steenbergen-2014-cheeger}.
Properly incorporating near-harmonic components into data analysis remains a challenge, 
but our framework provides a starting point.

\subsubsection*{Acknowledgments}
We thank Jean-Baptiste Seby for a careful reading of the manuscript and useful comments.
 
\bibliographystyle{siamplain}
\bibliography{./references.bib}

\newpage
\appendix

\section{Proof of \cref{thm:stoch_lifting}}
Recall the normalized Hodge 1-Laplacian is defined as 
\[
\bm{\mathcal{L}}_1 = \bm{D}_2\bm{B}_1^\top\bm{D}_1^{-1}\bm{B}_1^{} + \bm{B}_2^{}\bm{D}_3\bm{B}_2^\top \bm{D}_2^{-1},
\]
and our goal is to show that $-\bm{\mathcal{L}}_1\bm{V}^\top = 2\bm{V}^\top \bm{\widehat P}$ 
where $\bm{\widehat P}$ is defined in~\cref{thm:stoch_lifting}.
We will prove this result in three steps.
First, we show that 
\begin{equation}\label{eq:lifting_first_part}
    -(\bm{D}_2\bm{B}_1^\top\bm{D}_1^{-1}\bm{B}_1^{})\bm{V}^\top =  
    \bm{V}^\top\underbrace{\frac{1}{2}\left( \bm M_f \text{diag}(\bm{M}_f \bm{1})^{-1} + \bm M_b \text{diag}(\bm{M}_b \bm{1})^{-1}\right)}_{\bm P_{lower}},
\end{equation}
where 
\begin{align}
\bm M_f = \bm{\widehat{D}}_2(\liftB_1^-)^\top \liftB_1^+ \text{ and }\bm M_b = \bm{\widehat{D}}_2(\liftB_1^+)^\top \liftB_1^-
\end{align}
 are weighted lower adjacency matrices corresponding to forward and backward walks along the edges 
 (see~\cref{thm:stoch_lifting}), and $\bm{\widehat{D}}_2 = \text{diag}(\bm{D}_\mathrm{2},\bm{D}_\mathrm{2})$.

Second, we prove that
\begin{equation}\label{eq:lifting_second_part}
    -(\bm{B}_2^{}\bm{D}_3\bm{B}_2^\top \bm{D}_2^{-1})\bm{V}^\top = \bm{V}^\top\underbrace{\left(\widehat{\bm{A}}_u  \widehat{\bm{D}}_4^{-1}
    + \frac{1}{2} \begin{pmatrix} \bm I & \bm I \\ \bm I & \bm I \end{pmatrix} \widehat{\bm D}_5\right)}_{\bm P_{upper}},
\end{equation}
where $\widehat{\bm{A}}_u = \liftB_2^+(\liftB_2^-)^\top  + \liftB_2^-(\liftB_2^+)^\top$ is the matrix of 
upper adjacent connections, and $\widehat{\bm{D}}_4$ and $\widehat{\bm D}_5$ are as defined in~\cref{thm:stoch_lifting}.

Finally, we prove that ${\widehat{\bm P} = \frac{1}{2}(\bm P_{lower} +\bm P_{upper})}$ is indeed a stochastic matrix.

\subsection*{Preliminary results}
Before embarking on the proof, let us recall some useful facts and develop some lemmas.
For any matrix $\bm{M}$:
\begin{align}
\bm{M} &= \bm{M}^+ - \bm{M}^-, 
\qquad (-\bm{M})^- = \bm{M}^+, 
\qquad (-\bm{M})^+ = \bm{M}^-.
\end{align}
And from some algebra:
\begin{align}
& \bm{V}\bm{V}^\top = \bm{I}_{2n_1} - \bm{\Sigma}  \label{eq:vvt} \\
& \bm\Sigma \bm\Sigma = \bm{I}_{2n_1}  \label{eq:sigsig_id} \\
& \liftB_1 := \bm{B}_1\bm{V}^\top = \begin{pmatrix} \bm{B}_1 & -\bm{B}_1 \end{pmatrix} = \liftB_1^+ -\liftB_1^- \\
& \bm \liftB_1^- \bm \Sigma = \bm \liftB_1^+  \text{ and } \bm \liftB_1^- = \bm \liftB_1^+ \bm \Sigma \label{eq:lift1} \\
& \bm{B}_1\bm{V}^{\top} = \liftB_1^+\bm{V}\bm{V}^{\top} \label{eq:lift2}  \\
& \bm{B}_1 = \liftB_1^+ \bm V = -\liftB_1^- \bm V  \label{eq:lift3} \\
& \liftB_2 := \bm{V}\bm{B}_2 = \begin{pmatrix} \bm{B}_2 &  -\bm{B}_2\end{pmatrix}^{\top}  = \liftB_2^+ -\liftB_2^- \\
& \liftB_2^+ = \bm \Sigma \liftB_2^- \text{ and } \bm \Sigma \liftB_2^+ = \liftB_2^- \label{eq:lift4} \\
& \bm V \bm B_2 = \bm V \bm V^{\top} \liftB_2^+  \label{eq:lift5} \\
& \bm B_2 = \bm{V}^{\top} \liftB_2^+ = -\bm{V}^{\top} \liftB_2^-. \label{eq:lift6} 
\end{align}

Now, define the normalizing factors of the forward and backward walk using
lower adjacency relationships as
\begin{align*}
\bm{Q}_f = \text{diag}(\bm{1}^{\top} \bm{M}_f ) \text{ and } \bm{Q}_b = \text{diag}(\bm{1}^{\top} \bm{M}_b).
\end{align*}
\begin{lemma}\label{lem:app_lem1}
    $\bm{Q}_f \bm\Sigma = \bm\Sigma\bm{Q}_b$.
\end{lemma}
\begin{proof}
The diagonal entries of $\bm{Q}_f$ are given by the vector
\begin{align*}
    \bm q_f^{\top}
     := \bm{1}^\top\bm{M}_f &= \bm{1}^\top\bm{\widehat{D}}_2(\liftB_1^-)^\top \liftB_1^+  \\
    &= \bm{1}^\top\bm{\widehat{D}}_2\bm\Sigma\bm\Sigma(\liftB_1^-)^\top \liftB_1^+ \\
    &=\bm{1}^\top\bm{\widehat{D}}_2 \bm\Sigma (\liftB_1^-)^\top \liftB_1^+ \\    
    &=\bm{1}^\top\bm{\widehat{D}}_2 (\liftB_1^+)^\top \liftB_1^+ \\
    &=\bm{1}^\top \begin{pmatrix} \bm{D}_2 & \bm{0} \\ \bm{0} & \bm{D}_2 \end{pmatrix} 
    \begin{pmatrix} (\bm B_1^+)^{\top}\bm B_1^+ & (\bm B_1^+)^{\top}\bm B_1^-  \\
                                (\bm B_1^-)^{\top}\bm B_1^+ & (\bm B_1^-)^{\top}\bm B_1^-  \\
   \end{pmatrix} \\
   &= \bm{1}^\top \begin{pmatrix} 
   \bm{D}_2 \lvert \bm B_1 \rvert^{\top} \bm B_1^+  &
   \bm{D}_2 \lvert \bm B_1 \rvert^{\top} (\bm B_1^-)  \\ 
   \end{pmatrix}.
\end{align*}
The first line follows the definition of $\bm{Q}_f$;
the second line uses \cref{eq:sigsig_id};
the third line follows from $\bm 1^\top \widehat{\bm{D}}_2 \bm \Sigma =\bm 1^\top \widehat{\bm{D}}_2$;
the fourth line uses \cref{eq:lift1}; and the
fifth and sixth lines follows from expanding notation.
Similarly, the diagonal entries of $\bm{Q}_b$ are given by
\begin{align*}
\bm q_b^{\top} 
 := \bm{1}^\top\bm{M}_b 
&= \bm{1}^\top \bm{\widehat{D}}_2(\liftB_1^+)^\top \liftB_1^- \\
&= \bm{1}^\top \bm{\widehat{D}}_2 \bm\Sigma\bm\Sigma (\liftB_1^+)^\top \liftB_1^- \\
&= \bm{1}^\top \bm{\widehat{D}}_2 \bm\Sigma (\liftB_1^+)^\top \liftB_1^- \\
&=\bm{1}^\top\bm{\widehat{D}}_2(\liftB_1^-)^\top \liftB_1^- \\
&=\bm{1}^\top \begin{pmatrix} \bm{D}_2 & \bm{0} \\ \bm{0} & \bm{D}_2 \end{pmatrix} 
    \begin{pmatrix} (\bm B_1^-)^{\top}\bm B_1^- & (\bm B_1^-)^{\top}\bm B_1^+  \\
                                (\bm B_1^+)^{\top}\bm B_1^- & (\bm B_1^+)^{\top}\bm B_1^-  \\
   \end{pmatrix} \\
&= \bm{1}^\top \begin{pmatrix} 
   \bm{D}_2 \lvert \bm B_1 \rvert^{\top} \bm B_1^-  &
   \bm{D}_2 \lvert \bm B_1 \rvert^{\top} \bm B_1^+ \\
   \end{pmatrix} 
   = \bm{q}_f^{\top} \bm \Sigma.
\end{align*}
The lemma then follows by considering how the permutation $\bm \Sigma$ acts on the diagonal matrices $\bm{Q}_f$ and $\bm{Q}_b$.
\end{proof}

\begin{lemma}\label{lem:app_lem2}
    $\bm{D}_1^{-1}\liftB_1^{+} =  \frac{1}{2}\liftB_1^+\bm{Q}_f^{-1}$.
\end{lemma}
\begin{proof}
    By definition, $\liftB_1^+ = \begin{pmatrix} \bm{B}_1^+ & \bm{B}_1^-\end{pmatrix}$ 
    is an indicator matrix with a single entry equal to $1$ per column.
    More specifically, $\bm{B}_1^+$ picks out the target nodes of the oriented edges and $\bm{B}_1^-$ picks out the source-nodes:
    \begin{align*}
        (\bm{B}_1^+)_{[i],[j,k]} = 
        \begin{cases}
            1 & \text{if } i=k\\
            0 & \text{otherwise},
        \end{cases} \quad
        (\bm{B}_1^-)_{[i],[j,k]} = 
        \begin{cases}
            1 & \text{if } i=j\\
            0 & \text{otherwise}.
        \end{cases} 
    \end{align*}
    Recall that $\bm D_1 = 2 \cdot \text{diag} (\bm{a})$, where $\bm{a} = \lvert \bm{B}_1  \rvert \bm{D}_2 \bm{1}$. 
    Thus,
    \begin{align*}
        \left[ \bm{D}_1^{-1} \bm{B}_1^+ \right]_{[i],[j,k]} = 
        \begin{cases}
            \frac{1}{2a_{ii}} & \text{if } i=k\\
            0 & \text{otherwise},
        \end{cases} \quad
        \left[ \bm{D}_1^{-1}\bm{B}_1^- \right]_{[i],[j,k]} = 
        \begin{cases}
            \frac{1}{2a_{ii}} & \text{if } i=j\\
            0 & \text{otherwise}.
        \end{cases} 
    \end{align*}        
    From the proof of \cref{lem:app_lem1},
    \begin{equation*}
        \bm Q_f 
        = \text{diag} \begin{pmatrix} \bm Q_{f, 1} \\ \bm Q_{f, 2} \end{pmatrix}
        =  \text{diag} \begin{pmatrix} (\bm B_1^+)^{\top}\bm{a} \\ (\bm B_1^-)^{\top}\bm{a} \end{pmatrix}.
    \end{equation*}
    Row $[j, k]$ of $(\bm B_1^+)^{\top}$ equals $\bm e_k^{\top}$ and row $[j, k]$ of $(\bm B_1^-)^{\top}$ equals $\bm e_j^{\top}$.
    Thus, 
    \begin{align*}
        \left[ \frac{1}{2} \bm{B}_1^+ \bm Q_{f,1}^{-1} \right]_{[i],[j,k]} = 
        \begin{cases}
            \frac{1}{2a_{kk}} & \text{if } i=k\\
            0 & \text{otherwise},
        \end{cases} \quad
        \left[ \frac{1}{2} \bm{B}_1^- \bm Q_{f,2}^{-1} \right]_{[i],[j,k]} = 
        \begin{cases}
            \frac{1}{2a_{jj}} & \text{if } i=j\\
            0 & \text{otherwise.}
        \end{cases}
    \end{align*}
    Since the case statements hold when $i = k$ (first block) or $i = j$ (second block), we get the desired equality. 
\end{proof}

\subsection*{Proof of \cref{eq:lifting_first_part}}
We now prove the first part of \cref{thm:stoch_lifting}, i.e., \Cref{eq:lifting_first_part}.
\begin{proof}
    \begin{align*}
        -2[\bm{D}_2\bm{B}_1^\top\bm{D}_1^{-1}\bm{B}_1^{}]\bm{V}^\top 
        &= -2\bm{D}_2\bm{B}_1^\top\bm{D}_1^{-1}\liftB_1^{+}\bm{V}\bm{V}^\top 
        \quad \cref{eq:lift2} \\
        &=-\bm{D}_2\bm{B}_1^\top \liftB_1^+\bm{Q}_f^{-1} \bm{VV}^\top
        \quad (\cref{lem:app_lem2}) \\
        &=-\bm{D}_2\bm{B}_1^\top \liftB_1^+\bm{Q}_f^{-1} (\bm{I} -\bm\Sigma) 
        \quad \cref{eq:vvt} \\
        &= -\bm{D}_2\bm{B}_1^\top \liftB_1^+\bm{Q}_f^{-1} + \bm{D}_2\bm{B}_1^\top \liftB_1^+\bm{Q}_f^{-1}\bm\Sigma
        \quad (\text{expansion}) \\
        &= -\bm{D}_2\bm{B}_1^\top \liftB_1^+\bm{Q}_f^{-1} + \bm{D}_2\bm{B}_1^\top \liftB_1^-\bm \Sigma\bm{Q}_f^{-1}\bm\Sigma
        \quad \cref{eq:lift1} \\
        &= -\bm{D}_2\bm{B}_1^\top \liftB_1^+\bm{Q}_f^{-1} + \bm{D}_2\bm{B}_1^\top \liftB_1^-\bm{Q}_b^{-1}\bm \Sigma\bm\Sigma
        \quad (\cref{lem:app_lem1}) \\
        &= -\bm{D}_2\bm{B}_1^\top \liftB_1^+\bm{Q}_f^{-1} + \bm{D}_2\bm{B}_1^\top \liftB_1^-\bm{Q}_b^{-1}
        \quad \cref{eq:sigsig_id} \\
        &=  \bm{D}_2\bm{V}^\top(\liftB_1^-)^\top \liftB_1^+\bm{Q}_f^{-1} + \bm{D}_2\bm{V}^\top(\liftB_1^+)^\top \liftB_1^-\bm{Q}_b^{-1}
        \quad \cref{eq:lift3} \\
        &=\bm{V}^\top \bm{\widehat{D}}_2(\liftB_1^-)^\top \liftB_1^+\bm{Q}_f^{-1} + \bm{V}^\top \bm{\widehat{D}}_2(\liftB_1^+)^\top \liftB_1^-\bm{Q}_b^{-1}\\
        &= \bm{V}^\top \left( \bm M_f \bm Q_f^{-1} + \bm M_b \bm Q_b^{-1}\right)
        \quad (\text{by definition}).
    \end{align*}
    Note again that $\bm M_f \bm Q_f^{-1}$ and $\bm M_b \bm Q_b^{-1}$ are simply transition matrices of
    random walks on graph with weighted adjacency matrices $\bm M_f$ and $\bm M_b$; 
    accordingly, any convex combination is also a valid transition matrix.
\end{proof}

\subsection*{Proof of \cref{eq:lifting_second_part}}
Finally, we prove the second part of \cref{thm:stoch_lifting}, i.e., \Cref{eq:lifting_second_part}.
\begin{proof}
First, observe that
\begin{equation*}
    \bm{V}^\top\begin{pmatrix} \bm I & \bm I \\ \bm I & \bm I \end{pmatrix} \widehat{\bm D}_5 = \bm 0 \widehat{\bm D}_5 = \bm 0,
\end{equation*}
which implies that the contribution to our projection of the walks that correspond to edges without an upper-adjacent face is zero.
Hence, it suffices to show the following:
\begin{equation*}
    -(\bm{B}_2^{}\bm{D}_3\bm{B}_2^\top \bm{D}_2^{-1})\bm{V}^\top = \bm{V}^\top\widehat{\bm{A}}_u  \widehat{\bm{D}}_4^{-1}.
\end{equation*}
We then have that
    \begin{align*}
        \bm{V}^\top \widehat{\bm{A}}_u 
        &= \bm{V}^\top \left[\liftB_2^+(\liftB_2^-)^{\top} + \liftB_2^-(\liftB_2^+)^{\top}\right] 
        \quad \text{(definition of $\widehat{\bm{A}}_u$)} \\
        &= \bm{B}_2(\widehat{\bm{B}}_2^-)^\top - \bm{B}_2(\widehat{\bm{B}}_2^+)^\top 
        \quad \cref{eq:lift6}\\
        &= -\bm{B}_2\left[(\widehat{\bm{B}}_2^+)^\top - (\widehat{\bm{B}}_2^-)^\top\right] \\
        &= -\bm{B}_2(\widehat{\bm{B}}_2^+)^\top(\bm{I} - \bm\Sigma) 
        \quad \cref{eq:lift4} \\
        &= -\bm{B}_2(\widehat{\bm{B}}_2^+)^\top\bm V \bm V^{\top} 
        \quad \cref{eq:vvt} \\
        &= -\bm{B}_2\bm{B}_2^\top\bm{V}^\top
        \quad \cref{eq:lift5}.
    \end{align*}
    Since $\bm D_3$ is simply a scaling by $1/3$,
    \begin{align*}
        \bm{V}^\top \widehat{\bm{A}}_u \widehat{\bm{D}}_4^{-1}
        &=  -\bm{B}_2 \bm D_3 \bm{B}_2^\top \bm{V}^\top (3 \cdot \widehat{\bm{D}}_4^{-1}) \\
        &=  -\bm{B}_2 \bm D_3 \bm{B}_2^\top \bm{V}^\top \widehat{\bm{D}}_2^{-1} \\
        &=  -\bm{B}_2 \bm D_3 \bm{B}_2^\top  \bm{D}_2^{-1}\bm{V}^\top.
    \end{align*}
    In the second equality, we used the fact that if $\text{deg}([i, j]) = 0$, then the corresponding
    column of $\bm{B}_2 \bm D_3 \bm{B}_2^\top$ will be zero.
\end{proof}

\subsection*{Stochasticity of $\widehat{\bm P}$}
The matrix $\bm P_{\text{lower}}$ is column stochastic by construction. The last item we need to show
is that $\bm P_{\text{upper}}$ is also column stochastic so that we indeed have a stochastic lifting of
the normalized Hodge 1-Laplacian. Recall that
\begin{align*}
\bm P_{\text{upper}} &=  \widehat{\bm{A}}_u \widehat{\bm{D}}_4^{-1} +
\frac{1}{2} \begin{pmatrix} \bm I & \bm I \\ \bm I & \bm I \end{pmatrix} \widehat{\bm D}_5 \\
&=  \left[\liftB_2^+(\liftB_2^-)^{\top} + \liftB_2^-(\liftB_2^+)^{\top}\right] \widehat{\bm{D}}_4^{-1} +
\frac{1}{2} \begin{pmatrix} \bm I & \bm I \\ \bm I & \bm I \end{pmatrix} \widehat{\bm D}_5.
\end{align*}
When $\text{deg}([i, j]) = 0$, the corresponding column of $\widehat{\bm{A}}_u$ is zero and
$\widehat{\bm D}_5$ picks out a column of the stacked identity matrices, which is stochastic
when multiplied by $1 / 2$.
When $\text{deg}([i, j]) > 0$, the corresponding column of $\widehat{\bm{A}}_u$ has non-zero entries for exactly those edges that have a different orientation relative to each
co-face of $[i, j]$ (each co-face results in 3 entries, see \cref{eq:A_upper}). 
Thus, the column sum is exactly $3 \cdot \text{deg}([i, j])$, so scaling by $\widehat{\bm{D}}_4^{-1}$ makes the matrix stochastic.
\end{document}